\documentstyle[aps]{revtex}
\begin{document}
\title{Nonsingular Bianchi type I cosmological solutions\\ 
from the 1-loop superstring effective action
}
\author{Shinsuke {\sc Kawai}\footnote{E-mail:kawai@phys.h.kyoto-u.ac.jp}}
\address{Graduate School of Human and Environmental Studies, Kyoto University,
Kyoto 606-8501, Japan}
\author{Jiro {\sc Soda}\footnote{E-mail:jiro@phys.h.kyoto-u.ac.jp}}
\address{Department of Fundamental Sciences, FIHS, Kyoto University, Kyoto 
606-8501, Japan}
\date{Revised Jan 1999}
\maketitle
\begin{abstract}
 Nonsingular Bianchi type I solutions are found from the effective action 
with a superstring-motivated Gauss-Bonnet term. 
These anisotropic nonsingular solutions evolve from the asymptotic Minkowski
region, subsequently super-inflate, and then smoothly continue either to 
Kasner-type (expanding in two directions and shrinking in one direction) 
or to Friedmann-type (expanding in all directions) solutions. 
We also find a new kind of singularity which arises from the fact that the 
anisotropic expansion rates are a multiple-valued function of time.
The initial singularity in the isotropic limit of this model belongs to this 
new kind of singularity. 
In our analysis the anisotropic solutions are likely to be singular when the 
superinflation is steep. As for the cosmic no-hair conjecture, our results 
suggest that the kinetic-driven superinflation of our model does not 
isotropize the space-time. 
 
\end{abstract}
\twocolumn
\section{INTRODUCTION}
Stimulated by the developments of superstring theory, various 
cosmological solutions based on string theory have been proposed. 
Although the theory has not developed enough to present a unique history
of our universe in its earliest stage, one may now imagine that the
big bang is no longer just a point of enigma named the initial singularity but 
has a rich and complex structure. 

Among these string-based universe models, the most widely
studied would be the so-called pre-big-bang model\cite{pbb} (in the 
literature it is often called string cosmology). The remarkable aspect of 
this model is that it tries to explain the inflationary behavior of the early 
universe by introducing a polelike acceleration phase (superinflation) 
which is driven by the kinetic term of the dilaton. This superinflationary
branch of the solution has a dual relation named the scale factor duality with 
the 
Friedmann branch, which is the usual decelerating expansion of the universe.
The biggest problem arising in the pre-big-bang model is the difficulty of 
connecting the superinflationary branch and the Friedmann branch (termed the
graceful exit problem), and there are no-go theorems proved under some 
assumptions\cite{gexit}. 

Antoniadis, Rizos, and Tamvakis\cite{art94} proposed a nonsingular 
(i.e. free of the graceful exit problem) cosmological model by including the 
1-loop (genus) correction term in the low-energy string effective action.
This 1-loop
model is excellent in that it gives a simple example of the smooth transition
between the superinflationary branch and the Friedmann branch. This 
original 1-loop model involves dilaton and modulus fields, and there is a 
simplified version\cite{rt94} by Rizos and Tamvakis with a modulus field only. 
This metric-modulus system gives essentially the same nonsingular solution as 
the full metric-dilaton-modulus system, since the behavior of the solution 
mainly 
depends only on the modulus field. The existence of the nonsingular 
solution is analytically shown by Rizos and Tamvakis\cite{rt94} for the
Friedmann-Robertson-Walker metric, i.e., assuming the homogeneity and isotropy 
of the universe.
We studied  whether the nature of this nonsingular solution is affected if 
anisotropy is included. The purpose of this paper is to extend the solution 
of this metric-modulus system to include anisotropy and to observe the 
behavior of its solutions, particularly of nonsingular ones. 

The following sections of this paper are organized as follows.
In the Sec. II we briefly review the isotropic case studied by Rizos and
Tamvakis\cite{rt94}, and then derive the basic equations of motion from the 
effective action for the Bianchi type I metric. We also solve these equations 
analytically in the asymptotic region. 
In Sec. III we solve these equations numerically. 
We study the solutions through several cross sections of the parameter
space. The existence of nonsingular anisotropic 
solutions is shown, and the nature of the singularity is also examined.
Implications of our results are discussed in the last section.

\section{MODEL AND EQUATIONS OF MOTION}
We start with the action\footnote{
This action is also justifiable in more solid grounds. 
In the $N=4$ superstring compactifications (heterotic on $T^6$ or
type IIA on $K3\times T^2$) the function
$\xi(\sigma)$ given in Eq.(\ref{eqn:ssxi}) describes the exact $R^2$ couplings 
including nonperturbative effects\cite{hm96}.
These vacua exhibit an exact $S$ duality,
which is identical with the SL$(2,Z)$ modular invariance on $\sigma$.
} given by\cite{rt94}
\begin{equation}
{\cal S}=\int d^4x\sqrt{-g}\left\{\frac 12 R-\frac 12 (D\sigma)^2
 -\frac{\lambda}{16}\xi(\sigma) R_{GB}^2\right\},
\label{eqn:rt94act}
\end{equation}
which is essentially the same as the 1-loop-corrected 4-dimensional effective 
action of orbifold-compactified heterotic string\cite{art94},
\begin{eqnarray}
{\cal S}=\int d^4x\sqrt{-g}\left\{\frac 12 R-\frac 14(D\Phi)^2-\frac 34 
(D\sigma)^2\right.
 \nonumber\\
+\left.\frac 1{16} [\lambda_1 e^\Phi-\lambda_2\xi(\sigma)]R^2_{GB}\right\},
\label{eqn:ss1leAct}
\end{eqnarray}
except that the dilaton field $\Phi$ is neglected.
$R$, $\Phi$, and $\sigma$ are the Ricci scalar curvature, the dilaton, and
the modulus field, respectively. Our convention is $g_{\mu\nu}=(-,+,+,+)$, 
$R^\mu{}_{\alpha\nu\beta}=\Gamma^\mu{}_{\alpha\beta,\nu}+\cdots$,
$R_{\alpha\beta}=R^\mu{}_{\alpha\mu\beta}$, and $8\pi G=1$.
The Gauss-Bonnet curvature is defined as
$
R^2_{GB}=R^{\mu\nu\kappa\lambda}R_{\mu\nu\kappa\lambda}
 -4 R^{\mu\nu}R_{\mu\nu}+R^2,
$
and $\xi(\sigma)$ is a function determining the coupling of $\sigma$ and the 
geometry, written in terms of the Dedekind $\eta$ function as
\begin{eqnarray}
\xi(\sigma)&=&-\ln[2e^\sigma\eta^4 (i e^\sigma)]\nonumber\\
&=&-\ln 2-\sigma+\frac{\pi e^\sigma}{3}-4\sum_{n=1}^{\infty}\ln
(1-e^{-2n\pi e^\sigma}).
\label{eqn:ssxi}
\end{eqnarray}
This $\xi(\sigma)$, an even function of $\sigma$, has a global minimum at 
$\sigma=0$ and increases exponentially as $\sigma\rightarrow\pm\infty$. 
$\lambda_1$ is the 4-dimensional 
string coupling and takes a positive value. $\lambda_2$ is proportional to the
4-dimensional trace anomaly of the $N=2$ sector and determined by the number 
of chiral, vector, and spin-$\frac 32$ supermultiplets. It is important that 
$\lambda_2$ can take positive values, since nonsingular solutions arise only 
when $\lambda_2>0$.
In our simplified model (\ref{eqn:rt94act}), therefore, we assume $\lambda$ to 
be positive (in actual numerical calculations we set $\lambda=1$) and adopt
the form of the $\xi$ function (\ref{eqn:ssxi}).

\subsection{Isotropic solutions}
First, we review the homogeneous and isotropic case which is
discussed in \cite{rt94}. We neglect the spatial curvature and write 
the metric in the flat FRW form
\begin{equation}
ds^2=-N(t)^2dt^2+a(t)^2(dx^2+dy^2+dz^2).
\label{eqn:frwmetric}
\end{equation}
Variation of the action (\ref{eqn:rt94act}) with respect to the lapse $N$, 
the scale factor $a$, and the modulus field $\sigma$ gives three equations
of motion as
\begin{eqnarray}
&&\dot\sigma^2=6H^2 \left(1-\frac \lambda 2 H\dot\xi\right),
\label{eqn:bg1}\\
&&(2\dot H + 5H^2)\left(1-\frac \lambda 2 H \dot \xi\right)
 +H^2\left(1-\frac \lambda 2 \ddot\xi\right)=0,
\label{eqn:bg2}\\
&&\ddot\sigma+3H\dot\sigma+\frac {3\lambda}{2}
 (\dot H+H^2)H^2\frac{\partial\xi}{\partial\sigma}=0,
\label{eqn:bg3}
\end{eqnarray}
where $H$ is the Hubble parameter $\dot a/a$, the overdot means
derivative with respect to physical time $t$, and we have set $N(t)=1$.
As a result of the absence of scales [we are only considering spatially flat 
metric 
(\ref{eqn:frwmetric})] and the existence of the constraint (\ref{eqn:bg1}),
the solutions are completely determined by a couple of first order differential
equations for two variables $H$ and $\sigma$; that is, if we give values of
$H$ and $\sigma$ at some time $t$, the preceding and following evolutions of 
the
solution are automatically determined by these equations. 

Figure 1 shows the $H$-$\sigma$ phase diagram of the isotropic system solved 
with initial conditions $H>0$ and $\dot\sigma>0$. These solution flows are 
distinguished by only one degree of freedom (for example, the value of $H$ 
at some fixed $\sigma>0$). There are singular solutions and nonsingular 
solutions, and it is shown in \cite{rt94} that all flows in the $H>0$, 
$\sigma<0$ quarter-plane continue smoothly to $H>0$, $\sigma>0$ quarter-plane,
but some flows in $H>0$, $\sigma>0$ quarter-plane go into singularity and do 
not continue to the $\sigma<0$ region. It is also shown in \cite{rt94} that 
the 
signs of $H$ and $\dot\sigma$ are conserved throughout the evolution of the 
system. Since $\dot\sigma$ is always positive in Fig. 1, time flows from
left to right. We consider that our Friedmann universe corresponds to 
the ``future''
region ($\sigma>0$) in Fig. 1, and we regard the ``past'' region 
($\sigma<0$) with increasing Hubble parameter as a superinflation, which is 
expected to solve the shortcomings of the big-bang model.

\subsection{Equations of motion}
We now extend the above model to anisotropic Bianchi type I space-time.
We write the metric as
\begin{equation}
ds^2=-N(t)^2dt^2+e^{2\alpha(t)}dx^2+e^{2\beta(t)}dy^2+e^{2\gamma(t)}dz^2,
\end{equation}
and define the anisotropic expansion rates as
\begin{equation}
  p=\dot\alpha, \hspace{1cm}q=\dot\beta, \hspace{1cm}r=\dot\gamma.
\end{equation}
The average expansion rate, which coincides with the Hubble parameter $H$ in 
the isotropic limit, is  
\begin{equation}
H_{\mbox{\scriptsize avr}}=\frac 13(p+q+r).
\end{equation}
The equations of motion are obtained by 
variation of the action (\ref{eqn:rt94act}) with respect to $N$, $\alpha$,
$\beta$, $\gamma$, and $\sigma$, viz.,
\begin{eqnarray}
&&pq+qr+rp-\frac 12{\dot\sigma}^2
-\frac 32\lambda\frac{\partial\xi}{\partial\sigma}\dot\sigma pqr=0
\label{eqn:b1eom1},\\
&&\dot p=\frac{(CA-EB)G+(EF-A^2)H+(AB-FC)Q}\Delta
\label{eqn:b1eom2},\\
&&\dot q=\frac{(BC-DA)G+(BA-FC)H+(FD-B^2)Q}\Delta
\label{eqn:b1eom3},\\
&&\dot r=\frac{(DE-C^2)G+(AC-BE)H+(BC-AD)Q}\Delta
\label{eqn:b1eom4},\\
&&\ddot\sigma+(p+q+r)\dot\sigma\nonumber\\
&&\mbox{\hspace{1mm}}+\frac 12 \lambda
\frac{\partial\xi}{\partial\sigma}\left\{\dot p qr+p\dot q r+pq\dot r
+pqr(p+q+r)\right\}=0
\label{eqn:b1eom5},
\end{eqnarray}
where 
\begin{eqnarray}
A&=&1-\frac\lambda 2\frac{\partial\xi}{\partial\sigma}\dot\sigma p
+\frac {\lambda^2}4 \left(\frac{\partial\xi}{\partial\sigma}\right)^2p^2qr,\\
B&=&1-\frac\lambda 2\frac{\partial\xi}{\partial\sigma}\dot\sigma q
+\frac {\lambda^2}4 \left(\frac{\partial\xi}{\partial\sigma}\right)^2pq^2r,\\
C&=&1-\frac\lambda 2\frac{\partial\xi}{\partial\sigma}\dot\sigma r
+\frac {\lambda^2}4 \left(\frac{\partial\xi}{\partial\sigma}\right)^2pqr^2,\\
D&=&\frac {\lambda^2}4 \left(\frac{\partial\xi}
{\partial\sigma}\right)^2q^2r^2,\\
E&=&\frac {\lambda^2}4 \left(\frac{\partial\xi}
{\partial\sigma}\right)^2r^2p^2,\\
F&=&\frac {\lambda^2}4 \left(\frac{\partial\xi}
{\partial\sigma}\right)^2p^2q^2,\\
G&=&-p^2-q^2-pq-\frac 12{\dot\sigma}^2
+\frac\lambda 2\frac{\partial^2\xi}{\partial\sigma^2}{\dot\sigma}^2 pq
\nonumber\\
&&-\frac\lambda 2\frac{\partial\xi}{\partial\sigma}\dot\sigma pqr
-\frac\lambda 4\left(\frac{\partial\xi}
{\partial\sigma}\right)^2p^2q^2r(p+q+r),\\
H&=&-q^2-r^2-qr-\frac 12{\dot\sigma}^2
+\frac\lambda 2\frac{\partial^2\xi}{\partial\sigma^2}{\dot\sigma}^2 qr
\nonumber\\
&&-\frac\lambda 2\frac{\partial\xi}{\partial\sigma}\dot\sigma pqr
-\frac\lambda 4\left(\frac{\partial\xi}
{\partial\sigma}\right)^2pq^2r^2(p+q+r),\\
Q&=&-r^2-p^2-rp-\frac 12{\dot\sigma}^2
+\frac\lambda 2\frac{\partial^2\xi}{\partial\sigma^2}{\dot\sigma}^2 rp
\nonumber\\
&&-\frac\lambda 2\frac{\partial\xi}{\partial\sigma}\dot\sigma pqr
-\frac\lambda 4\left(\frac{\partial\xi}{\partial\sigma}\right)^2p^2qr^2(p+q+r),
\end{eqnarray}
and
\begin{equation}
\Delta=2 ABC+DEF-FC^2-DA^2-EB^2
\label{eqn:delta}.
\end{equation}
In the isotropic limit, Eqs. (\ref{eqn:b1eom1}) and (\ref{eqn:b1eom5}) 
become Eqs. (\ref{eqn:bg1}) and (\ref{eqn:bg3}) respectively, 
and Eqs. (\ref{eqn:b1eom2}), (\ref{eqn:b1eom3}), and (\ref{eqn:b1eom4}) 
reduce to one equation (\ref{eqn:bg2}). 
Similar to the isotropic case, once the values of $\sigma$, 
$p$, $q$, and $r$ are given at some time $t$, and $\dot\sigma$ is given 
according to the constraint (\ref{eqn:b1eom1}), the system evolves along the 
flow of the solution following Eqs. (\ref{eqn:b1eom2})$-$(\ref{eqn:b1eom5}). 
In our numerical analysis we 
integrated Eqs. (\ref{eqn:b1eom2})$-$(\ref{eqn:b1eom5}) with the
initial condition satisfying the constraint (\ref{eqn:b1eom1}). 
This constraint equation (\ref{eqn:b1eom1}) is also used to estimate the
numerical error.

There are five parameters ($\sigma$, $\dot\sigma$, $p$, $q$, and $r$)
and one constraint (\ref{eqn:b1eom1}); so the solution flows are drawn in a
4-dimensional space, for example, $\sigma$-$p$-$q$-$r$ space.
Since the aim of this paper is to examine the effect of anisotropy on the 
nonsingular solution, we observe the change of the solution as we 
increase the anisotropy of the metric. To facilitate this we introduce two 
parameters indicating the anisotropy of the metric,
viz.,
\begin{equation}
X=\frac{p-r}{p+q+r},\hspace{5mm} Y=\frac{q-r}{p+q+r}.
\label{eqn:defxy}
\end{equation}
In this notation, $(X,Y)=(0,0)$ corresponds to the isotropic (FRW) metric,
and postulating $H_{\mbox{\scriptsize avr}}$ to be positive, the region 
surrounded by the 
lines $Y=2X+1$, $Y=\frac 12 X-\frac 12$, $Y=-X+1$ indicates the universe 
expanding in all directions (these lines are drawn in Figs. 3 and 6). 
The regions $Y<2X+1$, $Y>\frac 12 X-\frac 12$, 
$Y>-X+1$, etc., are the ``Kasner-like'' universe expanding in two directions 
and contracting in one direction. The regions $Y>2X+1$, 
$Y>\frac 12 X-\frac 12$, $Y>-X+1$, etc., describe the universe expanding in
one direction, contracting in two directions. The universe shrinking in all 
directions cannot be included if we use Eqs. (\ref{eqn:defxy}) and assume
$H_{\mbox{\scriptsize avr}}>0$. Instead of $\sigma$-$p$-$q$-$r$ we
 use $\sigma$-$H_{\mbox{\scriptsize avr}}$-$X$-$Y$ as four variables 
describing our system.

\subsection{Asymptotic solutions}
Before going to the numerical analysis, we study the asymptotic form of the 
solutions at $t\rightarrow\pm\infty$, $\vert\sigma\vert\rightarrow\infty$. 
We assume $|\sigma|$ to become large when $t\rightarrow\pm\infty$. 
The asymptotic form of the derivatives of the function $\xi$ is
\begin{eqnarray}
\frac{\partial \xi}{\partial \sigma}&\sim&\mbox{sgn}(\sigma)\frac\pi 3 
e^{\vert\sigma\vert}
\label{eqn:axi1},\\
\frac{\partial^2 \xi}{\partial \sigma^2}&\sim&\frac\pi 3e^{\vert\sigma\vert}
\label{eqn:axi2}.
\end{eqnarray}
If we assume the power-law ansatz for the expansion rates, the asymptotic form 
of the modulus has to be logarithmic in order to cancel the exponential
dependence of Eqs. (\ref{eqn:axi1}) and (\ref{eqn:axi2}). Thus we choose the 
following forms for the asymptotic solutions:
\begin{eqnarray}
p&\sim&\omega_1\vert t \vert^\rho,\\
q&\sim&\omega_2\vert t \vert^\rho,\\
r&\sim&\omega_3\vert t \vert^\rho,\\
\sigma&\sim&\sigma_0 +\omega_4\ln\vert t\vert.
\end{eqnarray}
Putting all these into Eqs. (\ref{eqn:b1eom1})$-$(\ref{eqn:b1eom5}), 
we obtain two possible asymptotic solutions:
\begin{eqnarray}
{\cal A}&:&\rho=-1,\omega_1+\omega_2+\omega_3=\mbox{sgn}(t),\nonumber\\
&&\omega_1{}^2+\omega_2{}^2+\omega_3{}^2+\omega_4{}^2=1,
\label{eqn:assola}.\\
{\cal B}&:&\rho=-2,\vert\omega_4\vert=5,\nonumber\\
&&\omega_1\omega_2\omega_3=-\mbox{sgn}(t)\frac{5\exp[-\sigma_0\mbox{sgn}
(\omega_4)]}{\lambda\pi}
\label{eqn:assolb}.
\end{eqnarray}
The solution ${\cal A}$ is obtained by balancing terms that do not 
include the 1-loop effect, i.e., the Gauss-Bonnet term, and thus describes the 
asymptotic behavior where the Gauss-Bonnet effect is negligible.  In the 
absence of the modulus field, ${\cal A}$ is nothing but the Bianchi type
I vacuum (Kasner) solution. 
The solution ${\cal B}$ is obtained by balancing the kinetic term of the
modulus and the Gauss-Bonnet term, and this solution corresponds to the phase 
where the Gauss-Bonnet term is important. 
Other possibilities of solutions are excluded as long as we impose
$\lambda>0$, which is, in the isotropic limit, a necessary
condition for the existence of the nonsingular solutions.

Following the isotropic case\cite{art94,rt94,maeda96}, 
we choose $\dot\sigma>0$ and assume the solution ${\cal A}$ in the future 
asymptotic region and ${\cal B}$ in the past asymptotic region. 
Then, in the region $t\rightarrow\infty$, the conditions (\ref{eqn:assola})
can be seen in the $\omega_1$-$\omega_2$-$\omega_3$ space as the 
cross section of the sphere of radius $\sqrt{1-\omega_4^2}$ centered at the 
origin with the $\omega_1+\omega_2+\omega_3=1$ plane. 
Depending on the asymptotic value of $\omega_4$, the asymptotic solutions of
$t\rightarrow\infty$ are categorized into two cases.

${\cal A}$1: $0\leq\omega_4<\sqrt\frac 12$.
The asymptotic solution is either expanding in all directions (Friedmann type) 
or expanding in two directions, shrinking in one direction (Kasner type).

${\cal A}$2: $\sqrt\frac 12\leq\omega_4\leq\sqrt\frac 23$. 
The asymptotic solution is Friedmann type only.

In terms of $X$ and $Y$ introduced in Eq. (\ref{eqn:defxy}), the asymptotic
solution is represented by a point on an arc of the ellipse 
$X^2+Y^2-XY=1-\frac 32\omega_4^2$.
Thus ${\cal A}$2 falls into the region inside the oval $X^2+Y^2-XY=\frac 14$, 
and ${\cal A}$1 is in the region between two ellipses $X^2+Y^2-XY=\frac 14$ 
and $X^2+Y^2-XY=1$. No asymptotic solutions exist in the region outside the 
ellipse $X^2+Y^2-XY=1$, where the constraint equation (\ref{eqn:b1eom1}) is not
satisfied. Therefore, at least in the far enough future region, it is
sufficient to examine solutions near the isotropic one.
The difference in our model from the Kasner (Bianchi type I vacuum) solution is
the existence of the field $\sigma$, which allows the existence of a
Friedmann-type solution in the future asymptotic region.

In the past asymptotic region, the condition (\ref{eqn:assolb}) indicates
$\omega_1\omega_2\omega_3>0$, i.e., one of the following.

${\cal B}$1: $\omega_1$, $\omega_2$, $\omega_3>0$.

${\cal B}$2: One of $\omega_i$ ($i=1,2,3$) is positive; two are negative.

This means that in the past asymptotic region the universe is either expanding 
in all directions or expanding in one direction, contracting in two directions.

\section{NUMERICAL RESULTS}
Once the anisotropy is included, the behavior of the solution deviates
substantially from the isotropic case. Figure 2a shows the average expansion
rate $H_{\mbox{\scriptsize avr}}=(p+q+r)/3$ versus the modulus $\sigma$ in the 
anisotropic (Bianchi type I) case, 
solved with initial anisotropy $X=0.1$, $Y=0.2$ at $\sigma=-10$. Unlike the
isotropic case (Fig. 1), some solution flows in the $\sigma<0$ region
do not continue smoothly to the $\sigma>0$ region, but terminate suddenly with
finite values of $\sigma$ and $H_{\mbox{\scriptsize avr}}$. At these unusual
singularities the time derivatives of $p$, $q$, and $r$ become infinite, 
although $p$, $q$,
and $r$ themselves stay finite (see Fig. 2b). 
This is because the value of $\Delta$, Eq. (\ref{eqn:delta}),
approaches zero, while the numerators of Eqs. (\ref{eqn:b1eom2}),
(\ref{eqn:b1eom3}), and (\ref{eqn:b1eom4}) stay finite. 
Thus, the function $\Delta$ plays an important role in the anisotropic case, 
and the regularity of the solutions depends largely on its behavior.

In the equations in our model there are four independent variables
$\sigma$-$H_{\mbox{\scriptsize avr}}$-$X$-$Y$. 
Since our interest is mainly in the vicinity of the isotropic solution, 
we examine the solutions which pass near the origin of the $X$-$Y$ plane, 
first at the $\sigma=-10$ section with several different values of 
$H_{\mbox{\scriptsize avr}}$ and next at the $\sigma=2.5$ section. 

\subsection{Solutions through the $\sigma=-10$ cross section}
It is helpful to consider the general behavior of $\Delta$ and the region
prohibited by the constraint equation before solving the equations numerically.
Rewriting $\Delta$, Eq. (\ref{eqn:delta}), and the constraint equation 
(\ref{eqn:b1eom1}) in terms of $\sigma$, $H_{\mbox{\scriptsize avr}}$, $X$, 
and $Y$, 
we can specify $\Delta>0$, $\Delta<0$, and prohibited regions on the $X$-$Y$ 
plane for fixed $\sigma$ and $H_{\mbox{\scriptsize avr}}$, which is shown in 
Fig. 3.
The dark-shaded region is the prohibited region, the light-shaded region is 
where $\Delta<0$, and the white region is where $\Delta>0$. Since $\Delta=0$
is not physically allowed [$\dot p$, $\dot q$, and $\dot r$ diverge from 
Eqs. (\ref{eqn:b1eom2}), (\ref{eqn:b1eom3}), and (\ref{eqn:b1eom4})], 
the solutions in the 
white region cannot go smoothly to the light-shaded region. Also indicated in
Fig. 3 are the lines $X+Y=1$, etc., discussed in Sec. II B.
We can see that the universe expanding in all directions always lies
in the $\Delta>0$ region. For larger $H_{\mbox{\scriptsize avr}}$ the 
prohibited region 
becomes thinner, and the white and light gray regions will be separated by the 
lines $X+Y=1$, etc.

Starting from the initial conditions $\sigma=-10$ and $H_{\mbox{\scriptsize 
avr}}=0.001$,
$0.005$, $0.01$, we solved the equations futureward and indicated the behavior
of the solutions in the $X$-$Y$ plane (Fig. 4). Because of the symmetry 
of the axes, we restrict the region to $X>0$, $Y>0$, and since we are 
not interested in the prohibited region, we only examined the vicinity of the 
origin.

The black region in Fig.4 is prohibited by the Hamiltonian constraint,
and the regions marked NS means nonsingular solutions. 
The difference between NSa and NSb is in their form in the future asymptotic 
region, where NSa has Friedmann-type (expanding in all directions) and NSb has 
Kasner-type (expanding in two directions and contracting in one direction) 
asymptotic solutions. Examples of NSa and NSb solutions are shown in the first 
and the second panels of Fig. 5. The solutions in the region marked S1 in 
Fig.4 lead to singularities where $\Delta\rightarrow0$ 
(we term this singularity type I) and these singular solutions are the same as 
those appearing in Fig.2. We divided the S1 solutions into two classes, S1a 
and S1b. 
S1a approaches the singularity as $p\dot p>0$, $q\dot q>0$, $r\dot r>0$, 
while S1b as $p\dot p<0$, $q\dot q<0$, $r\dot r<0$. 
This singularity, since it arises because $\Delta$ crosses zero, 
can be overpassed if we introduce a new ``time'' variable
\begin{equation}
d\tau=dt/\Delta
\label{eqn:tau}.
\end{equation} 
Using this $\tau$, two solutions S1a and S1b can be joined via the singularity,
which is shown in the third panel of Fig. 5 (S1a,b). 
The solution S1b, solved backwards in time, goes into another singularity which
has different property from the one between S1a and S1b. At this singularity, 
which we call type II,  $\Delta$ goes to $-\infty$, and at least one of the 
expansion rates ($q$ in the case of Fig. 5, S1a,b) diverges. We can say that 
this
solution (S1a and S1b joined together) comes regularly from $t=-\infty$,
turns back at type I singularity, and then goes backwards in time into a type
II singularity. Or we can also see this as two solutions, one coming from $t=
-\infty$ and the other from the type II singularity, ``pair-annihilate'' at 
one type I singularity.  
S2 is yet another solution, which comes from one type II singularity and
disappears into another type II singularity. 
As $H_{\mbox{\scriptsize avr}}$ becomes larger, the boundary between S1a and 
S1b ($\Delta =0$ line in Fig. 3) gets pushed to approach the line $X+Y=1$, and 
accordingly the nonsingular regions NSa and NSb become smaller. This is 
consistent with Fig. 2, which shows the existence of the upper limit of
$H_{\mbox{\scriptsize avr}}$ for the regular solution through the 
$\sigma<0$ region.

\subsection{Solutions through the $\sigma=2.5$ cross section}
As is expected from the isotropic case discussed in the previous section, 
the solutions through the $\sigma>0$ cross section are quite different from 
those
through the $\sigma<0$ section. In Fig. 6 we show the prohibited region 
(dark shaded), $\Delta>0$ region (white), and $\Delta<0$ region (light shaded) 
in the $X$-$Y$ plane, with $\sigma$ fixed to $2.5$. The elliptic allowed region
of small $H_{\mbox{\scriptsize avr}}$
is the one discussed in relation to the future asymptotic form of the
solution, which is expressed as $X^2+Y^2-XY=1$. As 
$H_{\mbox{\scriptsize avr}}$ takes 
large values the $\Delta=0$ contour takes complicated forms, and the region
connected to the isotropic solution becomes small.

Figure 7 shows the solutions passing through the $X$-$Y$ plane of the 
$\sigma=2.5$ cross section, and the time evolution of each type is shown 
in Fig. 8.
$H_{\mbox{\scriptsize avr}}$ is chosen to be 0.01, 0.05, and 0.1. 
As $H_{\mbox{\scriptsize avr}}$ increases the nonsingular region becomes 
smaller, 
and for $H_{\mbox{\scriptsize avr}}$ larger than 0.1 the nonsingular 
solution completely
disappears from the $X$-$Y$ plane. All the singularities appearing in Fig.7 
are type I, and these singular solutions can be extended further by using 
$\tau$ defined by Eq. (\ref{eqn:tau}). Just like the S1a and S1b solutions 
in the
previous subsection, S1c and S1d, extended beyond the type I singularity, 
turn back futureward and then go into the type II singularity. The only
difference between these and S1a,b is the direction of time, and the former
can be regarded as the ``pair creation'' of cosmological solutions, while the 
latter is the ``pair annihilation.'' 

One of the nontrivial results of our analysis, and
what makes this model very different from ordinary universe models, is that
the ``initial singularity'' in the isotropic limit is categorized into the type
I singularity (see the third panel of Fig. 7 and compare S1d and isotropic
solutions in Fig. 8). This means that the singular solutions in the model 
proposed by Antoniadis, Rizos, and Tamvakis\cite{art94} or Rizos and Tamvakis
\cite{rt94} will, if small anisotropy is included, terminate suddenly at finite
past with finite Hubble parameter or, if extended using $\tau$, turn back 
towards the future. 

All nonsingular solutions in Figs. 7 and 8 continue to the asymptotic
solutions expanding in all directions in the past asymptotic region. That is, 
the asymptotic form ${\cal B}$1 (discussed in Sec. II C) can be reached from 
$\sigma=2.5$ but ${\cal B}$2 cannot. 
There exist solutions having the past asymptotic form ${\cal B}$2. 
For example, solutions through the outer white ($\Delta>0$) region in the 
third panel of Fig. 3 behave as ${\cal B}$2 in the $t\rightarrow -\infty$ 
region. These solutions, however, go into singularities between $\sigma=-10$ 
and $\sigma=2.5$, and do not appear in Fig. 7 or 8.

\subsection{Nature of the singularity}
In our numerical analysis there appear two types of singularities, which we
called type I and type II. We discuss the nature of these singularities 
briefly.

At type I singularities the expansion rates ($p,q,r$) stay finite whereas the
time derivatives of them diverge. These situations happen when
$p,q,r\sim \vert t-t_{sing}\vert ^\rho$
with $0<\rho<1$. This is actually the case, and can be verified by analyzing 
solutions near the singularity. 
Although the type I singularity ($\Delta=0$) is a physical one, 
the equations of motions can be integrated regularly by using the new ``time'' 
parameter $\tau$, Eq. (\ref{eqn:tau}), through the type I singularity. 
At the singularity $p,q,r$ ``turn around'' (see the solutions S1 in Figs. 5 
and 8), meaning that $p$, $q$, and $r$ are multiple-valued function of $t$.
Since $dt=\Delta d\tau=0$ at the singularity, the solutions are tangential to 
$t=t_{sing}$. 
Also, in order that the solutions change the chronological direction, the 
leading power of $p,q,r$ in the expansion of $t$ near the singularity must be 
even. Thus we can express $t$ using $p,q,r$ as
\begin{eqnarray}
t&-&t_{sing}\nonumber\\
&=&t_{p,2l}(p-p_{sing})^{2l}+t_{p,2l+1}(p-p_{sing})^{2l+1}+\cdots\nonumber\\
&=&t_{q,2m}(q-q_{sing})^{2m}+t_{q,2m+1}(q-q_{sing})^{2m+1}+\cdots\nonumber\\
&=&t_{r,2n}(r-r_{sing})^{2n}+t_{r,2n+1}(r-r_{sing})^{2n+1}+\cdots,
\end{eqnarray}
with $l,m,n$ being positive integers. By solving these with respect to 
$p$, $q$, and $r$, we have
\begin{eqnarray}
p&=&p_{sing}+p_1(t-t_{sing})^{1/{2l}}+\cdots\\
q&=&q_{sing}+q_1(t-t_{sing})^{1/{2m}}+\cdots\\
r&=&r_{sing}+r_1(t-t_{sing})^{1/{2n}}+\cdots.
\end{eqnarray}
Thus, $0<1/2l, 1/2m, 1/2n<1$, and 
$\dot p=(p_1/2l)(t-t_{sing})^{1/2l-1}+\cdots $,
etc., will diverge. 
The behavior of $\dot \sigma$ is similar to that of $p$, $q$, and
$r$. In our numerical calculations (S1 of Figs. 5 and 8), $l,m,n$ take the 
values $l=m=n=1$, which is the most generic case.

In the vicinity of other types of singularities, analytic expressions of the 
solutions are obtained by assuming power-law behavior of the scale
factors and the regularity of the modulus field, just as in the isotropic case
\cite{art94}. Because of the anisotropy, there are 3 cases of singular 
solutions other than the type I.
\begin{eqnarray}
{\cal C}1&:&p\sim p_1/t,q\sim q_0, r\sim r_0, \sigma\sim\sigma_0,
\dot\sigma\sim\sigma_1,\ddot\sigma\sim\sigma_2\nonumber,\\
&&p_1=1, q_0+r_0-\frac 32 \lambda\left.\frac{\partial\xi}{\partial\sigma}
\right|_{\sigma_0}\sigma_1q_0r_0=0
\label{eqn:assolc1},\\
{\cal C}2&:&p\sim p_1/t,q\sim q_1/t, r\sim r_0, \sigma\sim\sigma_0,
\dot\sigma\sim\sigma_1,\ddot\sigma\sim\sigma_2\nonumber,\\
&&p_1=q_1=1, 1-\frac 32 \lambda\left.\frac{\partial\xi}{\partial\sigma}
\right|_{\sigma_0}\sigma_1r_0=0
\label{eqn:assolc2},\\
{\cal C}3&:&p\sim p_1/t,q\sim q_1/t, r\sim r_1/t, \sigma\sim\sigma_0,
\dot\sigma\sim\sigma_1 t,\ddot\sigma\sim\sigma_1\nonumber,\\
&&p_1=q_1=r_1=1, 1-\frac 12 \lambda\left.\frac{\partial\xi}{\partial\sigma}
\right|_{\sigma_0}\sigma_1=0
\label{eqn:assolc3},
\end{eqnarray}
where we have chosen the origin of $t$ at the singularity. 
${\cal C}1$ is the case where only one of the three expansion rates ($p,q,r$) 
is singular, ${\cal C}2$ two, and ${\cal C}3$ all three. 
The solution ${\cal C}1$ agrees well with our numerical results (S2 of Fig. 5).
Since $p_1,q_1,r_1=1$
(if not zero), the divergent behavior is determined by the sign of
$t$; i.e., if a solution goes into a singularity futureward
($t\rightarrow -0$), then $p\rightarrow-\infty$, etc., and if pastward
($t\rightarrow +0$), $p\rightarrow+\infty$, etc.  
Putting Eqs. (\ref{eqn:assolc1})$-$(\ref{eqn:assolc3}) into 
Eq. (\ref{eqn:delta}),
we have $\Delta\rightarrow-\infty$ for ${\cal C}1$ and ${\cal C}2$ but
$\Delta\rightarrow +0$ for ${\cal C}3$. Therefore, according to our definition
of the singularities ($\Delta\rightarrow 0$ for type I and $\Delta\rightarrow
-\infty$ for type II), ${\cal C}1$ and ${\cal C}2$ will be categorized into
type II and ${\cal C}3$ will be categorized into type I. 
In the isotropic case ($p=q=r$), we can show that $\Delta>0$ is always 
satisfied, and ${\cal C}3$ can be seen as a type I singularity ``pushed 
towards the infinity.'' 

All of these singularities, both types I and II, are physical singularities.
This can be shown by putting $p,q,r$ and $\dot p,\dot q,\dot r$ into the 
curvature scalar
$R=2(\dot p+p^2+\dot q+q^2+\dot r+r^2+pq+qr+rp)$.

\subsection{Summary of numerical results}
We extended the nonsingular universe model proposed by Rizos and Tamvakis
\cite{rt94} to include anisotropy, and examined solutions in the
vicinity the of isotropic solution in both $\sigma<0$ and $\sigma>0$ regions.
We found both nonsingular solutions and singular solutions.
Nonsingular solutions inhabit the region where the anisotropy is 
small, and they evolve from the past asymptotic region, superinflate,
and then lead either to Friedmann-type or to Kasner-type solutions in the
future.  
Singularities appearing in our analysis are classified into two types, namely, 
type I and type II.
The type I singularity corresponds to the crossing of  $\Delta=0$, 
and $\dot p$, $\dot q$, and $\dot r$ diverge, 
while $p$, $q$, $r$, and $\sigma$, $\dot\sigma$ stay finite.
At the type II singularity, on the other hand, $\Delta$ tends to $-\infty$, and
$p$, $q$, and $r$ will diverge. 
The evolution of the singular solutions is characterized by the behavior of 
$\Delta$. At the origin of the $X$-$Y$ plane (isotropic solution) $\Delta$
is always positive regardless of the values of $\sigma$ or 
$H_{\mbox{\scriptsize avr}}$, 
and as anisotropy increases there appear $\Delta<0$ regions or regions
prohibited by the constraint equation (\ref{eqn:b1eom1}), which is shown in 
Figs. 3 and 6. There are three types of singular solutions appearing in our 
analysis (if two branches connected by a type I singularity are counted as one 
solution). The first type of singular solution is marked by the crossing of 
the $\Delta=0$ line, which we termed the type I singularity, in the future. 
This includes S1a and S1b in Figs. 4 and 5, and if we continue the 
solution 
beyond $\Delta=0$ by changing the variable, this singular solution can be seen 
as a pair annihilation of the $\Delta>0$ and $\Delta<0$ branches of solutions. 
The $\Delta>0$ branch continues from the infinite past, 
while the $\Delta<0$ branch 
leads to a type II singularity at the finite past.
The second singular solution is very similar to the first one, except it 
crosses the
$\Delta=0$ line in the past. This solution, examples of which are S1c and S1d 
of Figs.7 and 8, can be regarded as a pair creation of two branches. 
The singular solution in the isotropic model\cite{art94,rt94} is a special 
case of this second singular solution.
The third singular solution lies always in the $\Delta<0$ region and never 
crosses the
$\Delta=0$ line, i.e., includes no type I singularity. This solution is born in
the type II singularity, and disappears into the type II singularity (see S2 of
Figs. 4 and 5).
 
\section{CONCLUSION}
 In this paper we presented anisotropic nonsingular cosmological solutions 
derived from the 1-loop effective action of the heterotic string. We found
nonsingular solutions which evolve from the infinite past asymptotic region, 
superinflate, and then continue either to Friedmann-type or Kasner-type 
solutions. The singular solutions of moderate anisotropy are classified into
three types, and involved in these solutions are two types of singularities:
one corresponds to $\Delta=0$ and the other to $\Delta=-\infty$. The initial 
singularity in the isotropic limit is a special case of the $\Delta=0$ 
singularity.

Violation of the energy conditions, 
which is necessary to avoid the singularity, is achieved by the 
existence of a Gauss-Bonnet term coupled to a modulus field. This can be
confirmed by using the asymptotic forms (\ref{eqn:assola}) and 
(\ref{eqn:assolb}) for
nonsingular solutions. We define the effective energy density and effective
pressure as
\begin{eqnarray}
\epsilon&:=&-G^0{}_0=pq+qr+rp,\\
p_1&:=&G^1{}_1=-(\dot q+\dot r+q^2+r^2+qr),\\
p_2&:=&G^2{}_2=-(\dot r+\dot p+r^2+p^2+rp),\\
p_3&:=&G^3{}_3=-(\dot p+\dot q+p^2+q^2+pq).
\end{eqnarray}
Assuming the asymptotic solution ${\cal A}$ in the region $t\rightarrow\infty$
and that $\dot\sigma$ is always positive (in our numerical analysis 
$\dot\sigma$ keeps its sign except the singular solution S2), the effective 
energy density and pressure behave in the future asymptotic region as
$\epsilon$, $p_i\sim \frac 12 \omega_4^2\vert t\vert ^{-2}$ ($i=1,2,3$).
Thus, the weak energy condition $\epsilon+p_i\sim\omega_4^2\vert t\vert^{-2}>0$
and the strong energy condition $\epsilon+\sum p_i\sim 2\omega_4^2\vert t
\vert^{-2}>0$ are satisfied. In the past asymptotic region, on the other hand, 
the asymptotic solution becomes ${\cal B}$ in which the Gauss-Bonnet term is 
dominant. Then $\epsilon\sim (\omega_1\omega_2+\omega_2\omega_3
+\omega_3\omega_1)\vert t\vert ^{-4}$ 
and $p_1\sim-2(\omega_2+\omega_3)\vert t\vert^{-3}
-(\omega_2^2+\omega_3^2+\omega_2\omega_3)\vert t\vert^{-4}$, etc. 
If the weak energy condition is satisfied, 
$\epsilon+p_i>0$ for $i=1,2,3$, and so 
$3\epsilon+\sum p_i\sim-4(\omega_1+\omega_2+\omega_3)\vert t\vert^{-3}$
must be positive.
If the strong energy condition is satisfied,
$\epsilon+\sum p_i\sim-4(\omega_1+\omega_2+\omega_3)\vert t\vert^{-3}$ 
must be positive. 
Neither of them is possible as long as
$H_{\mbox{\scriptsize avr}}
\sim(\omega_1+\omega_2+\omega_3)\vert t\vert^{-2}/3>0$.
Therefore, weak and strong energy conditions are violated in the past 
asymptotic region.

One of the biggest advantages of our model is that it includes a rather
long period of (super)inflationary stage in a natural form. 
However, our result, which admits the evolution of an almost-isotropic 
superinflating solution into a Kasner-type anisotropic solution 
(see the panel ``NSb'' of Fig. 5 for example),
suggests that the superinflation in our model does not isotropize the 
space-time. Our recent study\cite{kss98} of the cosmological perturbation for
the homogeneous and isotropic background shows the existence of exponential 
growth in graviton-mode perturbation during the superinflationary stage. 
Together with this we can conclude that the cosmic no-hair hypothesis
\cite{cnohair} does not hold in this superinflationary model. 
We are interested in whether this result is common to all kinetic-driven 
superinflation.

\acknowledgements
 
This work of J.S. is supported by the Grant-in-Aid for Scientific 
Research No. 10740118.


\clearpage
{\Large Figures}
\input epsf.sty
\begin{figure}
\epsfxsize=9cm
\epsfysize=6cm
\epsffile{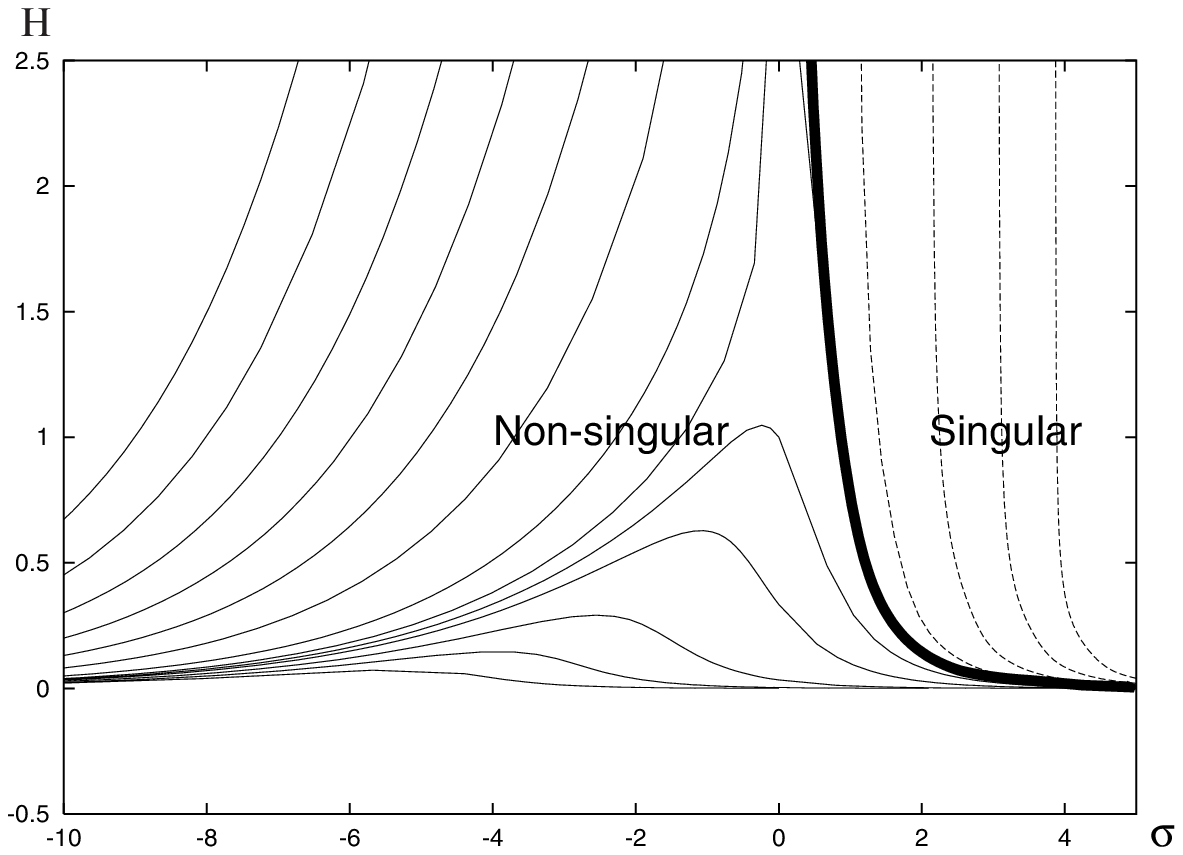}
\caption{The $H$-$\sigma$ phase diagram of isotropic solutions ($\lambda=1$,
$H>0$, $\dot\sigma>0$). Time flows from left to right since $\dot\sigma>0$. 
The nonsingular solutions are plotted with a solid line, and singular solutions
with a dotted line. The bold line is a critical solution marking the border 
of singular and nonsingular solutions. All solution flows in the $H>0$, 
$\sigma<0$
quarter-plane continue smoothly to the $H>0$, $\sigma>0$ quarter-plane. In the
$H>0$, $\sigma>0$ quarter-plane, however,  only the flows below the critical
solution continue to the $\sigma<0$ region.}
\end{figure}
\newpage
\begin{figure}
\epsfxsize=9cm
\epsfysize=12cm
\epsffile{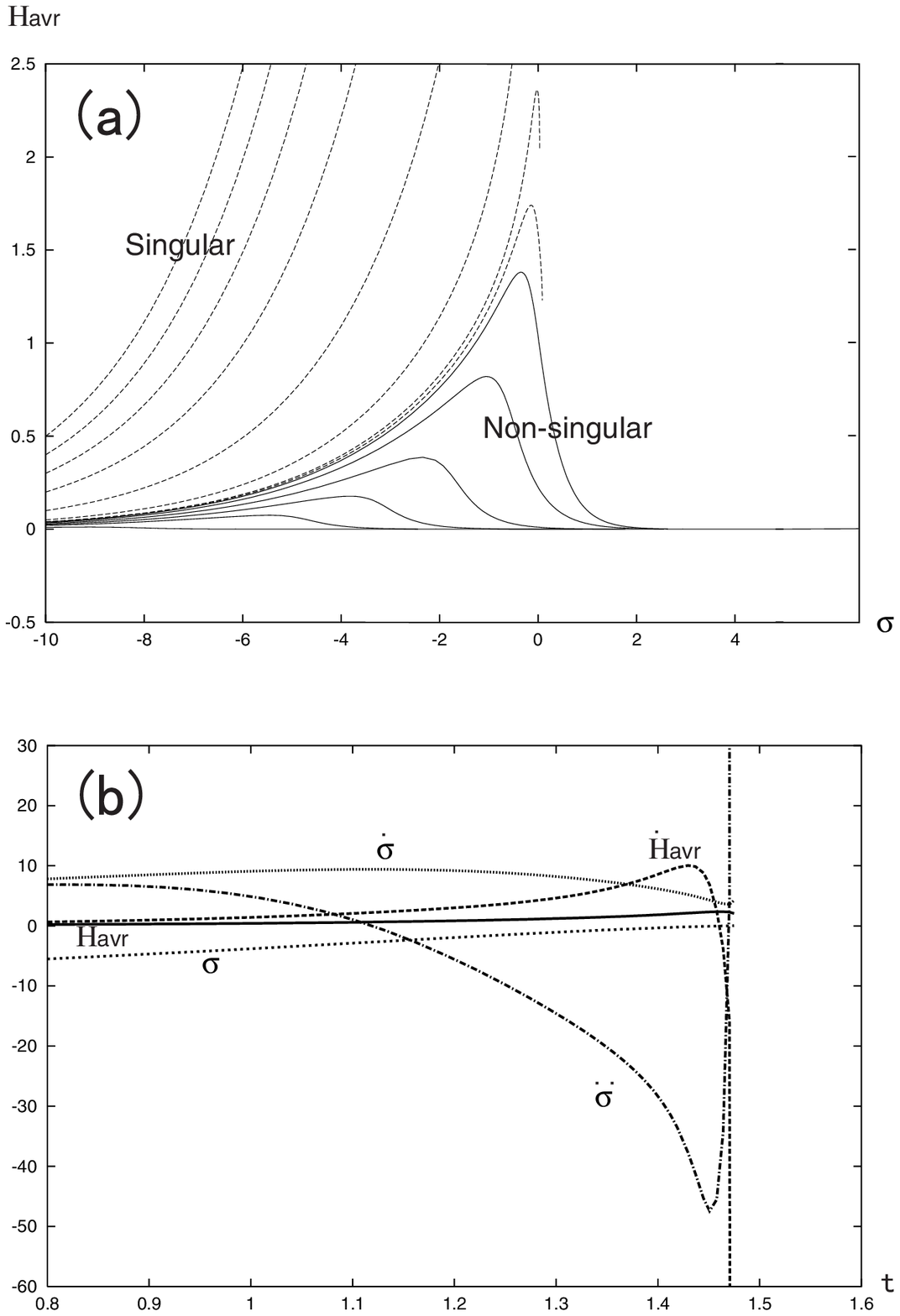}
\caption{(a) The average expansion rate in the anisotropic case. 
The equations are solved from $\sigma=-10$, where the initial anisotropy is 
fixed as $X=0.1$, $Y=0.2$. For large initial $H_{\mbox{\scriptsize avr}}$, 
there appear singularities with which the solution flows terminate suddenly, 
keeping $H_{\mbox{\scriptsize avr}}$ and $\sigma$ finite.
(b) The behavior of $H_{\mbox{\scriptsize avr}}$, 
$\dot H_{\mbox{\scriptsize avr}}$, $\sigma$, $\dot\sigma$, and 
$\ddot\sigma$ in a singular solution appearing in (a).
Initial conditions are the same as in (a) except the initial 
$H_{\mbox{\scriptsize avr}}$ is set to 0.04. 
We can see $\dot H_{\mbox{\scriptsize avr}}$ and $\ddot\sigma$ diverge, but
$H_{\mbox{\scriptsize avr}}$, $\sigma$, and $\dot\sigma$ stay finite.}
\end{figure}
\newpage
\begin{center}
\begin{figure}
$H_{\mbox{\scriptsize avr}}=0.001, \hspace{2mm}\sigma=-10$\hspace{12mm}
\epsfxsize=6cm
\epsfysize=6cm
\epsffile{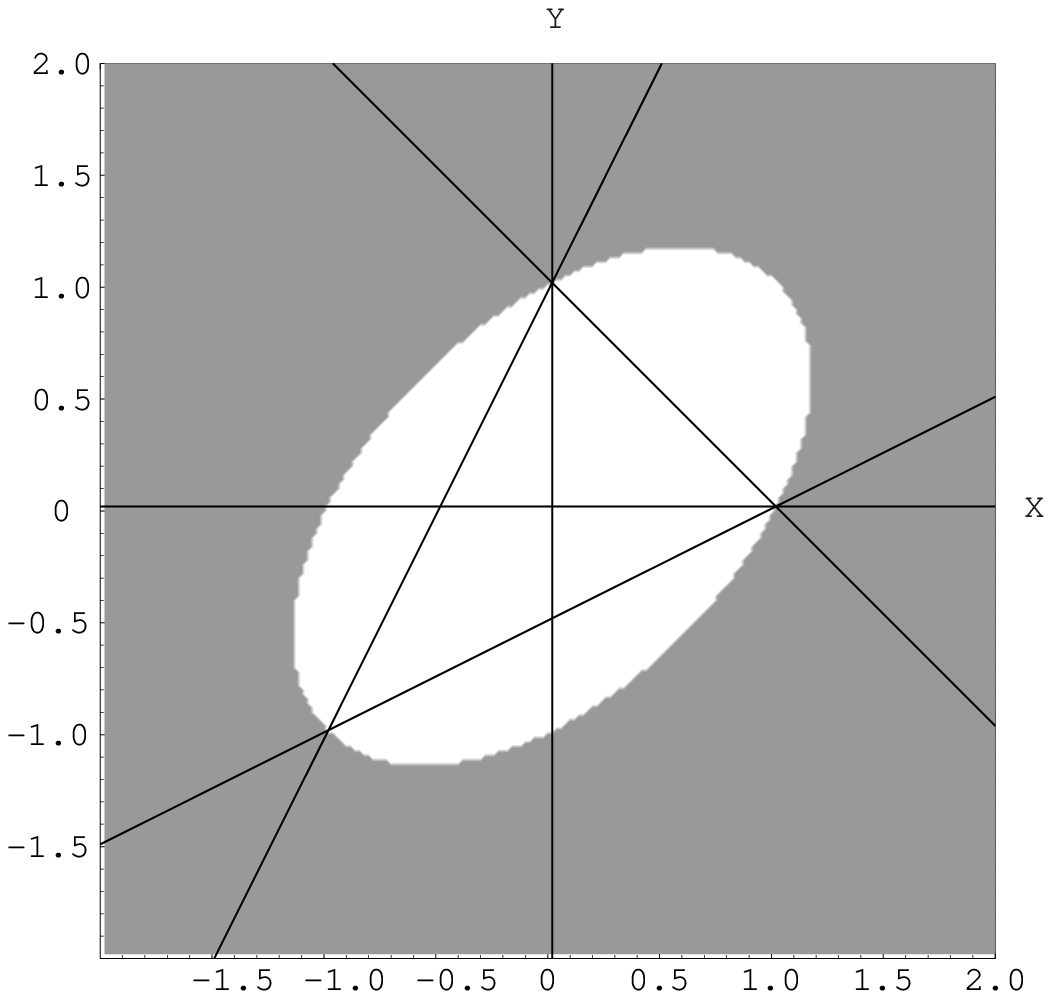}

$H_{\mbox{\scriptsize avr}}=0.005, \hspace{2mm}\sigma=-10$\hspace{12mm}
\epsfxsize=6cm
\epsfysize=6cm
\epsffile{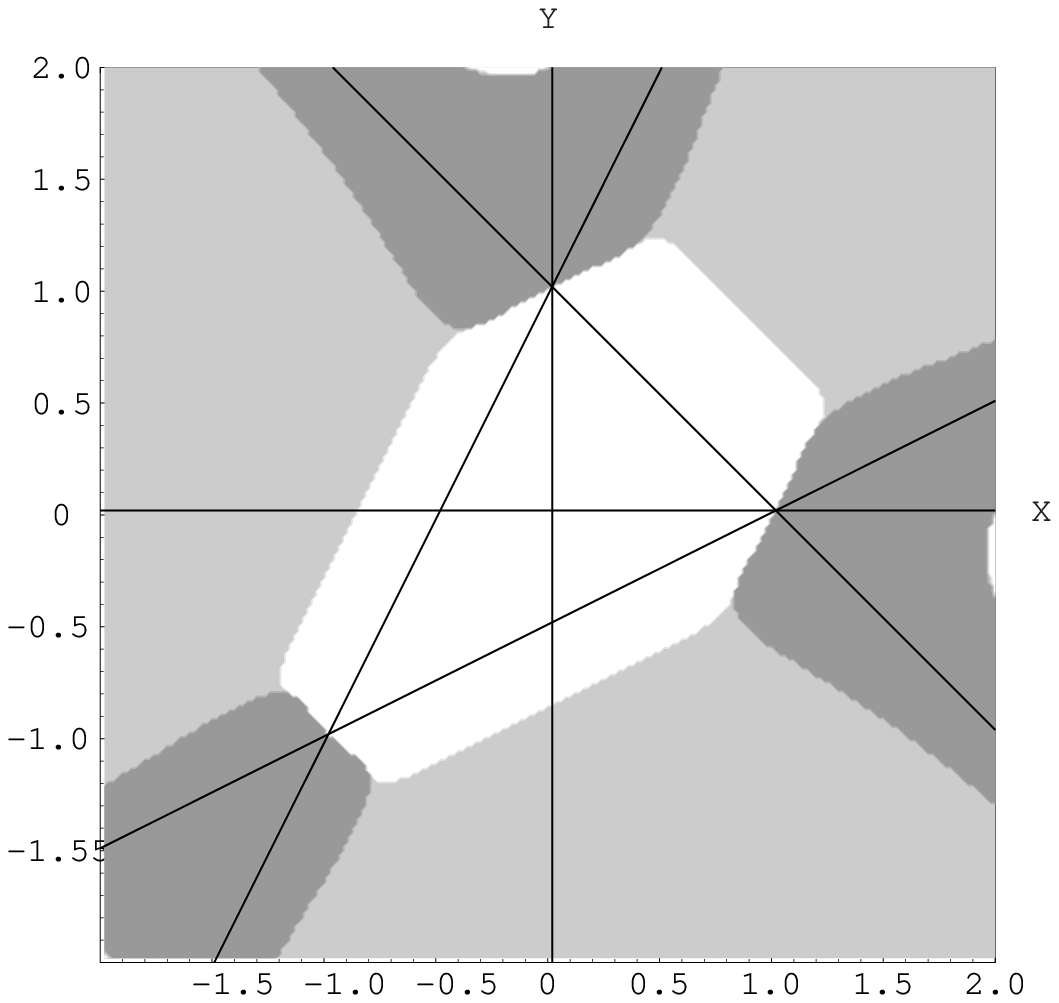}

$H_{\mbox{\scriptsize avr}}=0.01, \hspace{2mm}\sigma=-10$\hspace{12mm}
\epsfxsize=6cm
\epsfysize=6cm
\epsffile{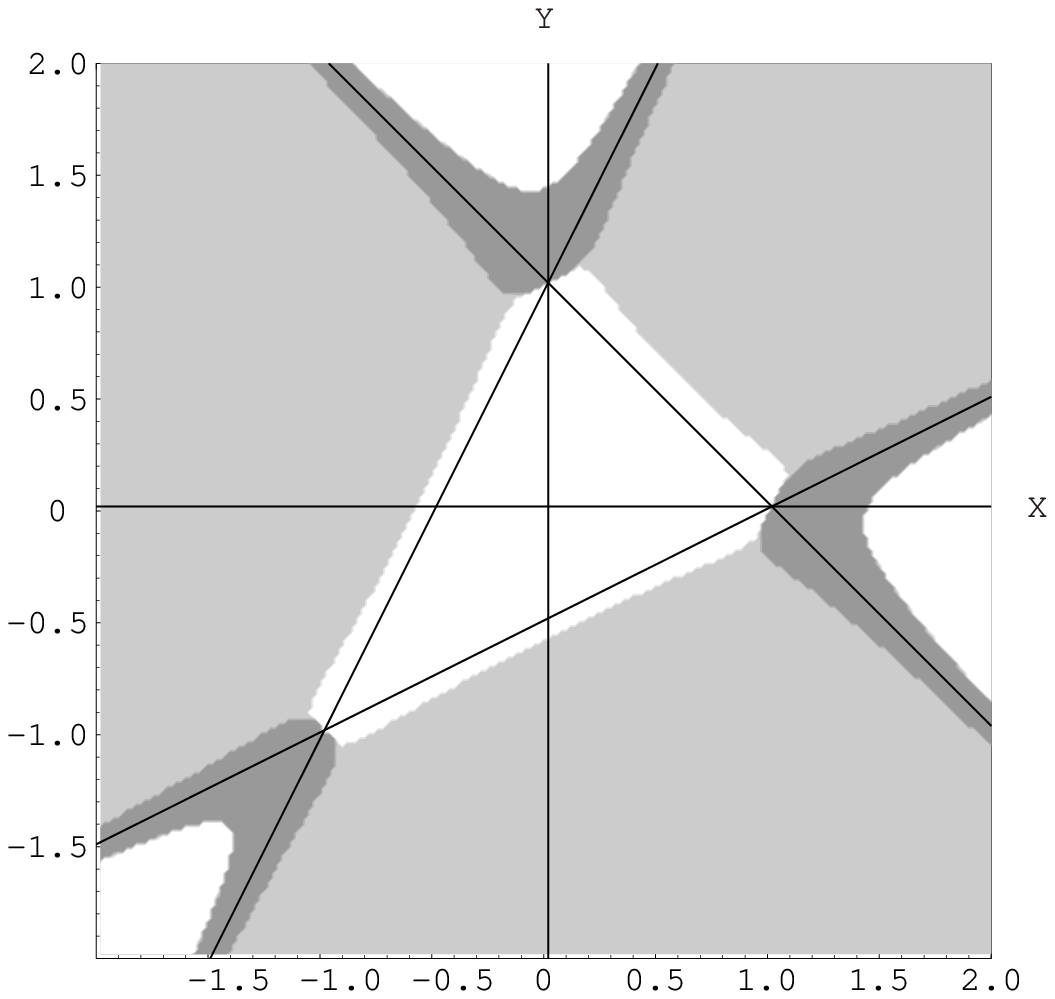}
\caption{Constraint and $\Delta$ on the $\sigma=-10$ section of the 
$X$-$Y$ plane.
$H_{\mbox{\scriptsize avr}}$ is 0.001, 0.005, 0.01 from above.
$\Delta>0$, $\Delta<0$, and excluded regions are indicated by white, 
light-shaded, and dark-shaded areas, respectively. 
In the dark gray region the Hamiltonian constraint (\ref{eqn:b1eom1}) 
is not satisfied, and $\Delta$ cannot be defined.  
Cosmological solutions inhabit the white and light gray regions, 
and those in each region are separated by singularities
since $p$, $q$, $r$ become infinite when $\Delta=0$ [see Eqs.
(\ref{eqn:b1eom2}), (\ref{eqn:b1eom3}), and (\ref{eqn:b1eom4})].}
\end{figure}
\end{center}
\onecolumn
\begin{figure}
$H_{\mbox{\scriptsize avr}}=0.001$, \hspace{2mm}$\sigma=-10$\hspace{30mm}
$H_{\mbox{\scriptsize avr}}=0.005$, \hspace{2mm}$\sigma=-10$\hspace{30mm}
$H_{\mbox{\scriptsize avr}}=0.01$, \hspace{2mm}$\sigma=-10$
\epsfxsize=6cm
\epsfysize=6cm
\epsffile{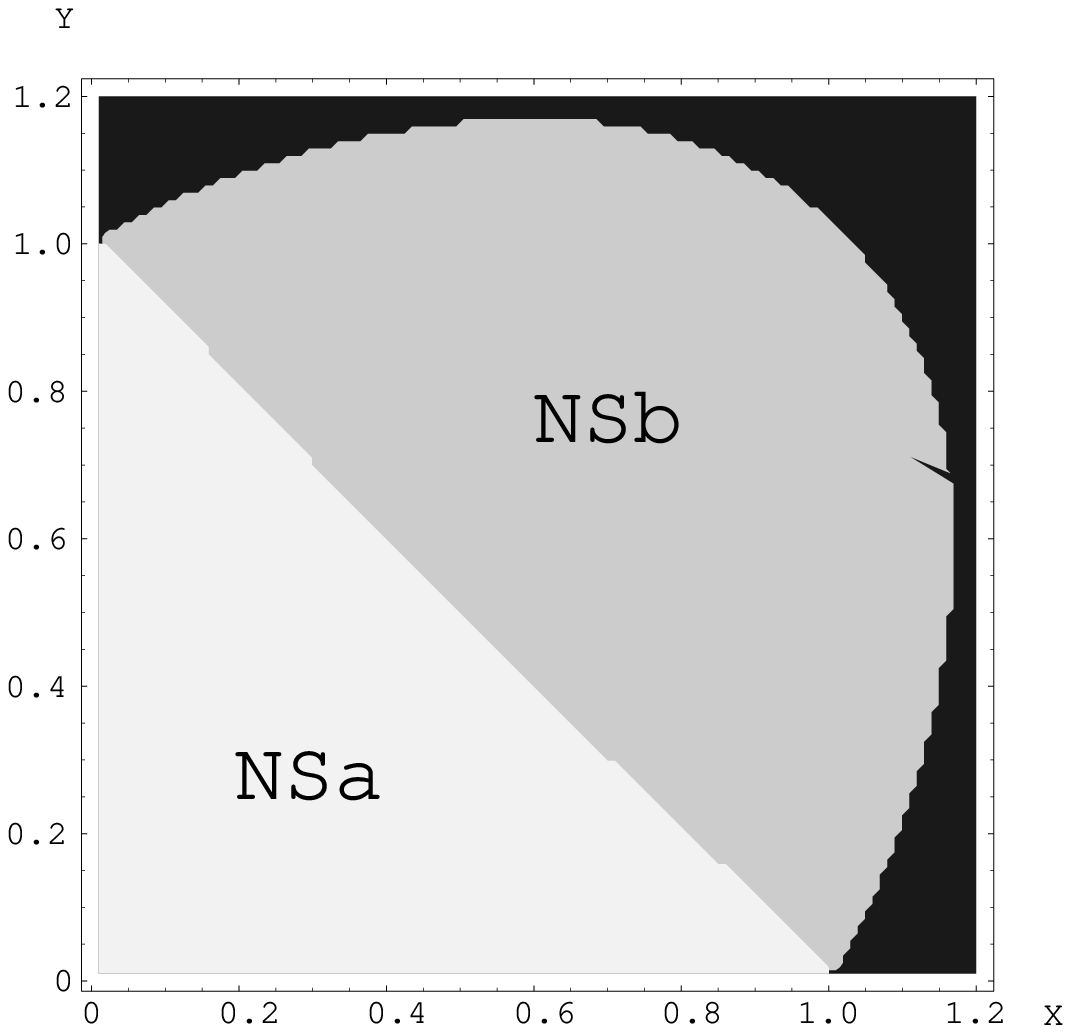}
\epsfxsize=6cm
\epsfysize=6cm
\epsffile{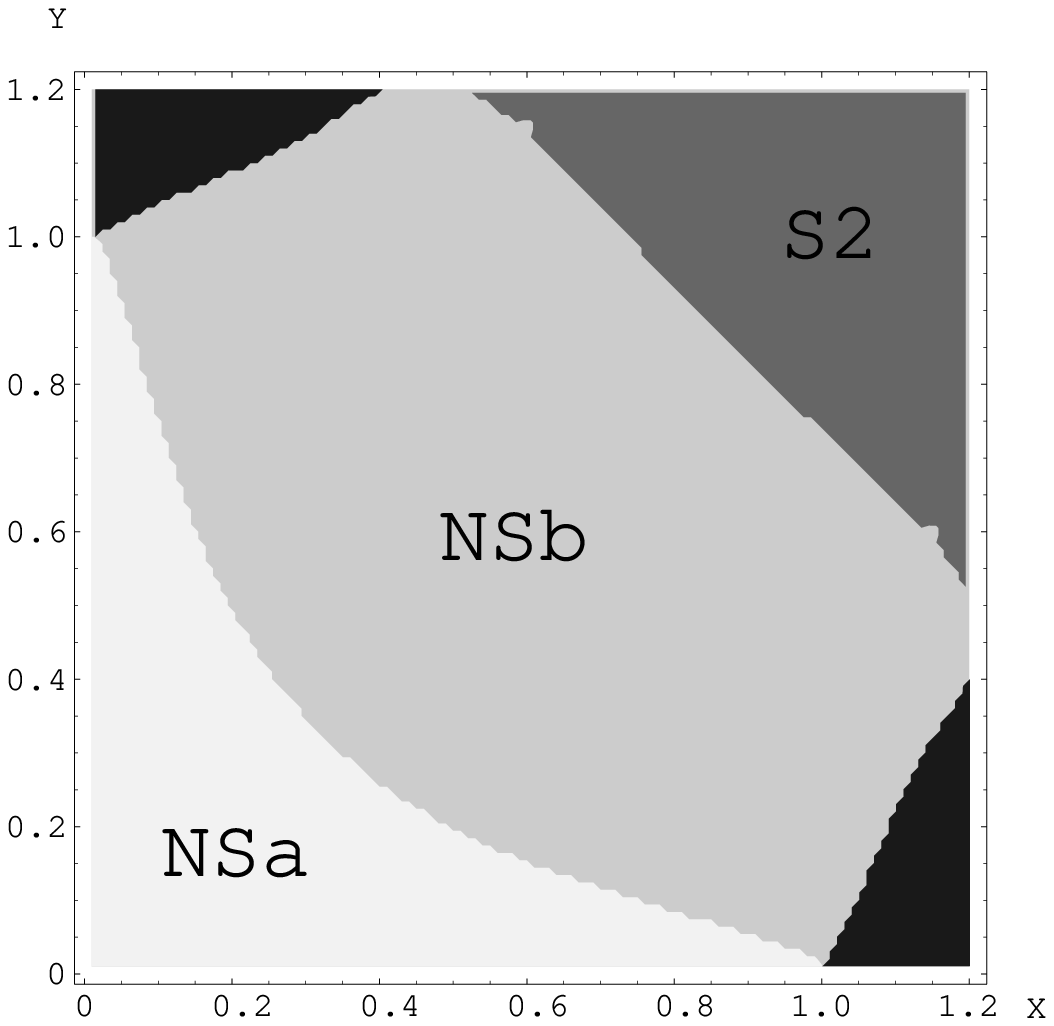}
\epsfxsize=6cm
\epsfysize=6cm
\epsffile{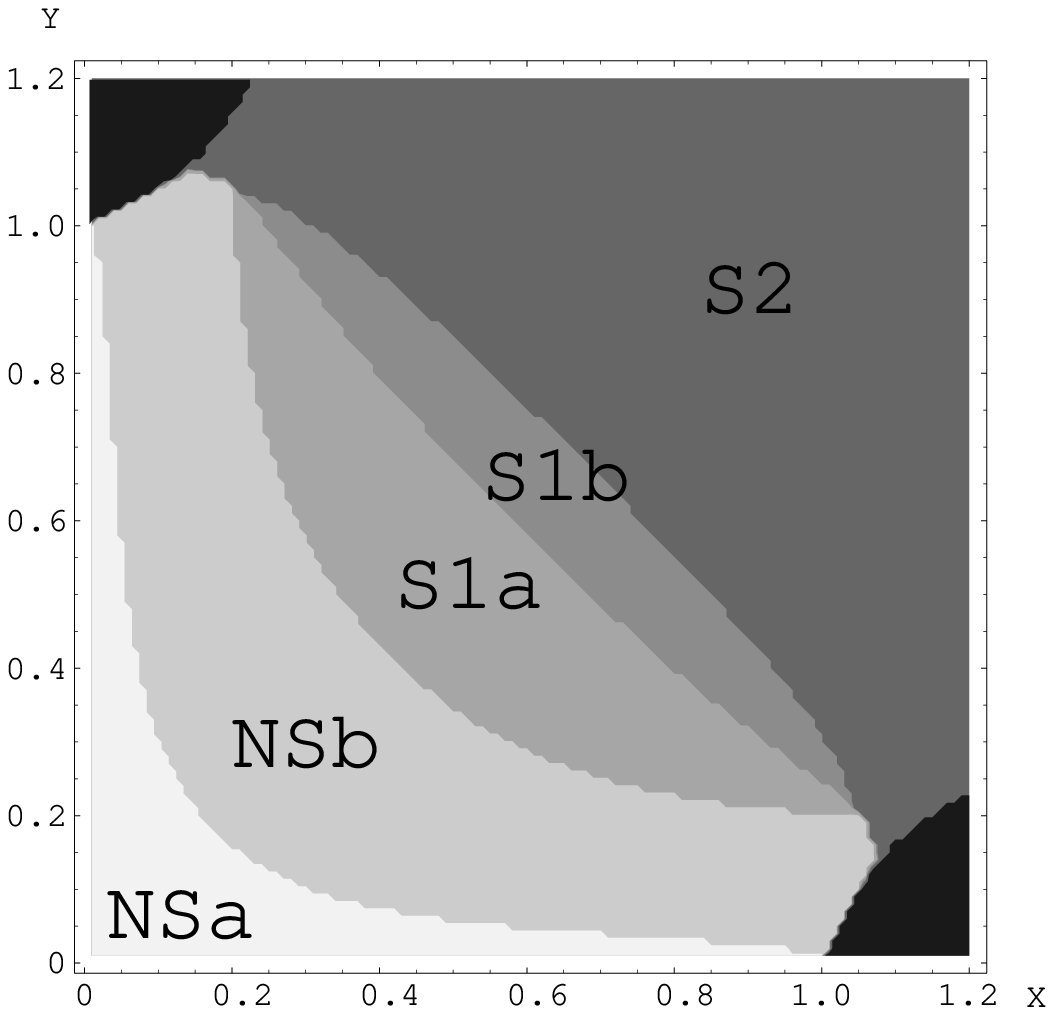}
\caption{Solutions through the $\sigma=-10$ cross section. 
$H_{\mbox{\scriptsize avr}}$ is 0.001,
0.005, 0.01 from the left. NS means nonsingular. NSa leads to a Friedmann-type
solution (expanding in all directions) and NSb leads to a Kasner-type solution 
(expanding in two directions and shrinking in one direction) in the future 
asymptotic region. S1 means it leads to a singularity where $\Delta\rightarrow
0$. S1a is the solution whose behavior near such a singularity is $p\dot p>0$, 
$q\dot q>0$, $r\dot r>0$, 
while S1b behaves as $p\dot p<0$, $q\dot q<0$, $r\dot r<0$, 
near the singularity. 
S2 leads to a singularity where $\Delta\rightarrow -\infty$.}
\end{figure}

\begin{figure}
\hspace{10mm}NSa\hspace{75mm}NSb
\epsfxsize=8cm
\epsfysize=5cm
\epsffile{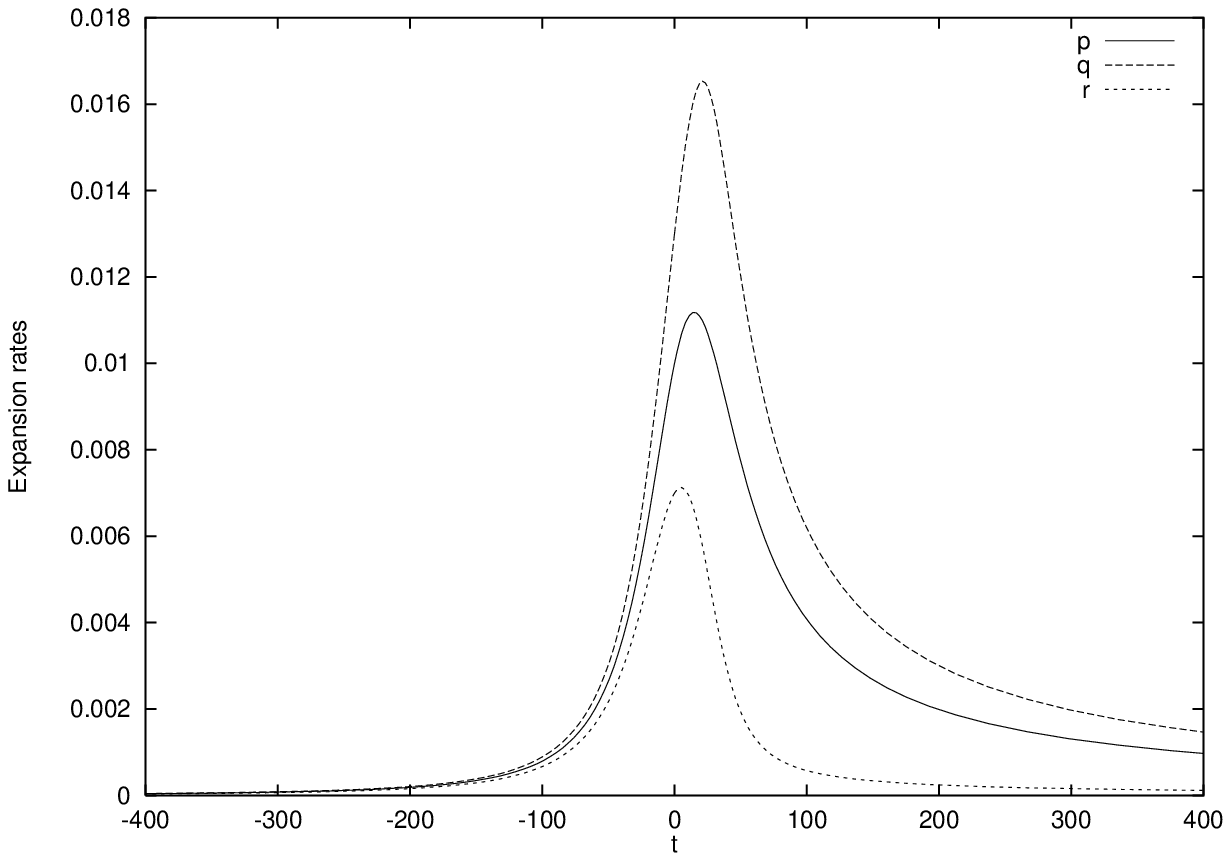}
\epsfxsize=8cm
\epsfysize=5cm
\epsffile{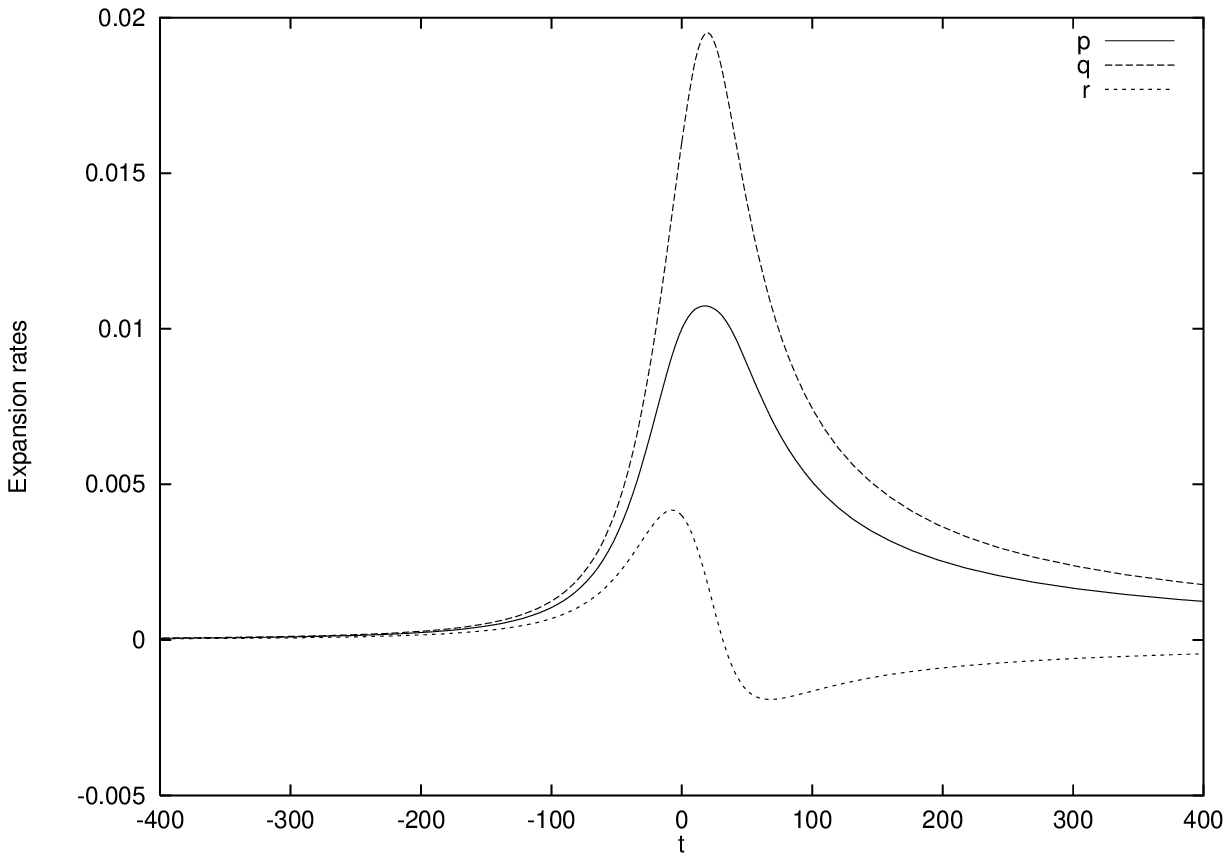}

\hspace{10mm}S1a,b\hspace{73mm}S2
\epsfxsize=8cm
\epsfysize=5cm
\epsffile{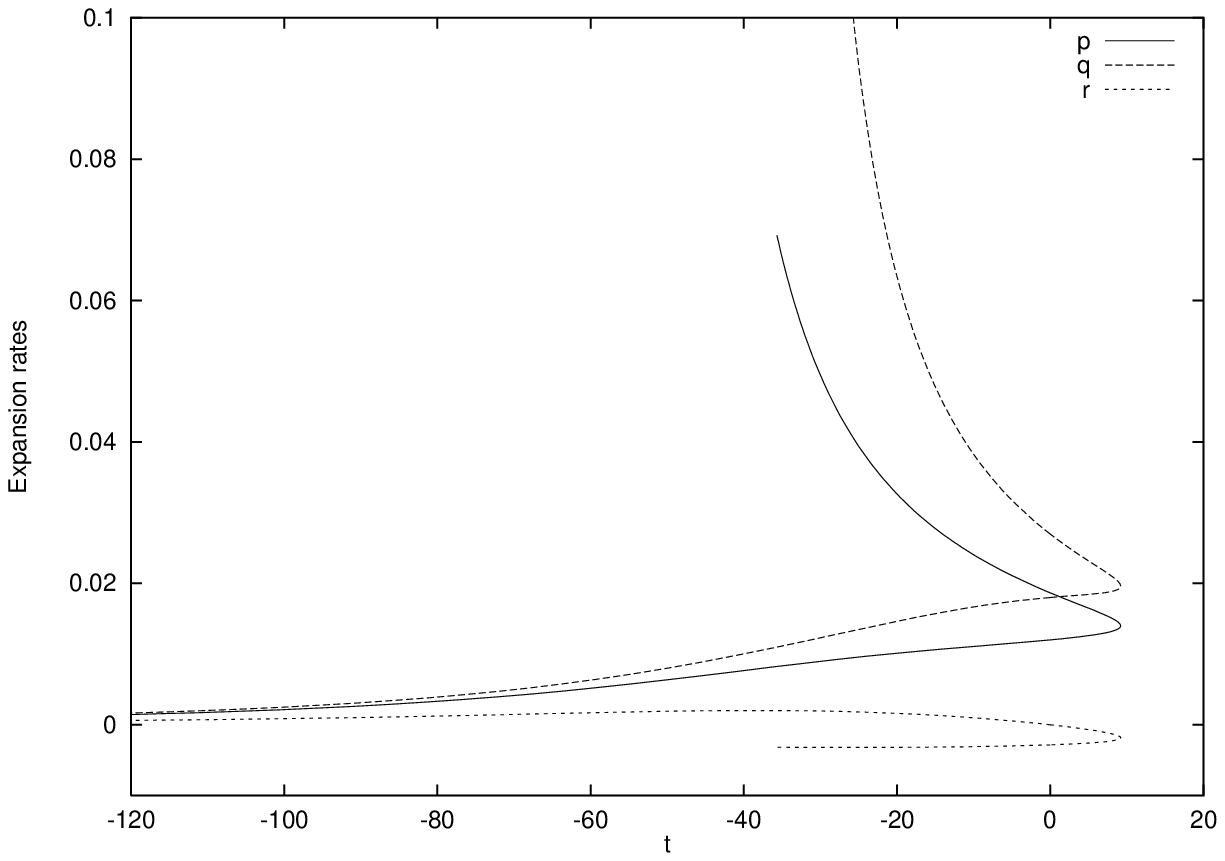}
\epsfxsize=8cm
\epsfysize=5cm
\epsffile{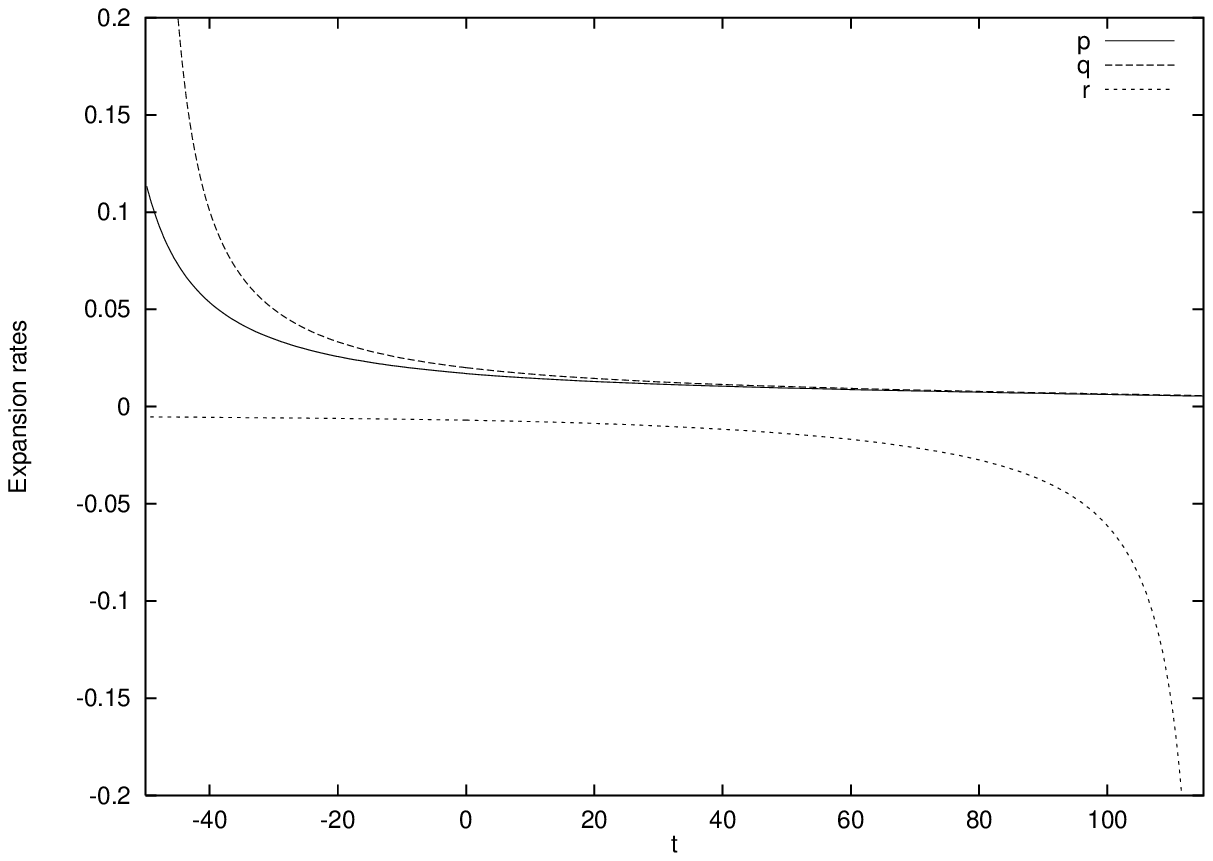}
\caption{Behavior of solutions appearing in Fig. 4. $t=0$ is the 
time when $\sigma=-10$. These are solutions through the points 
$(X,Y)=(0.1,0.2)$, $(0.2,0.4)$, $(0.4,0.6)$, and $(0.8,0.9)$, respectively,
in the $H_{\mbox{\scriptsize avr}}=0.01$, $\sigma=-10$ plane.}
\end{figure}
\twocolumn
\begin{center}
\begin{figure}
$H_{\mbox{\scriptsize avr}}=0.1$, \hspace{2mm}$\sigma=2.5$
\epsfxsize=6cm
\epsfysize=6cm
\epsffile{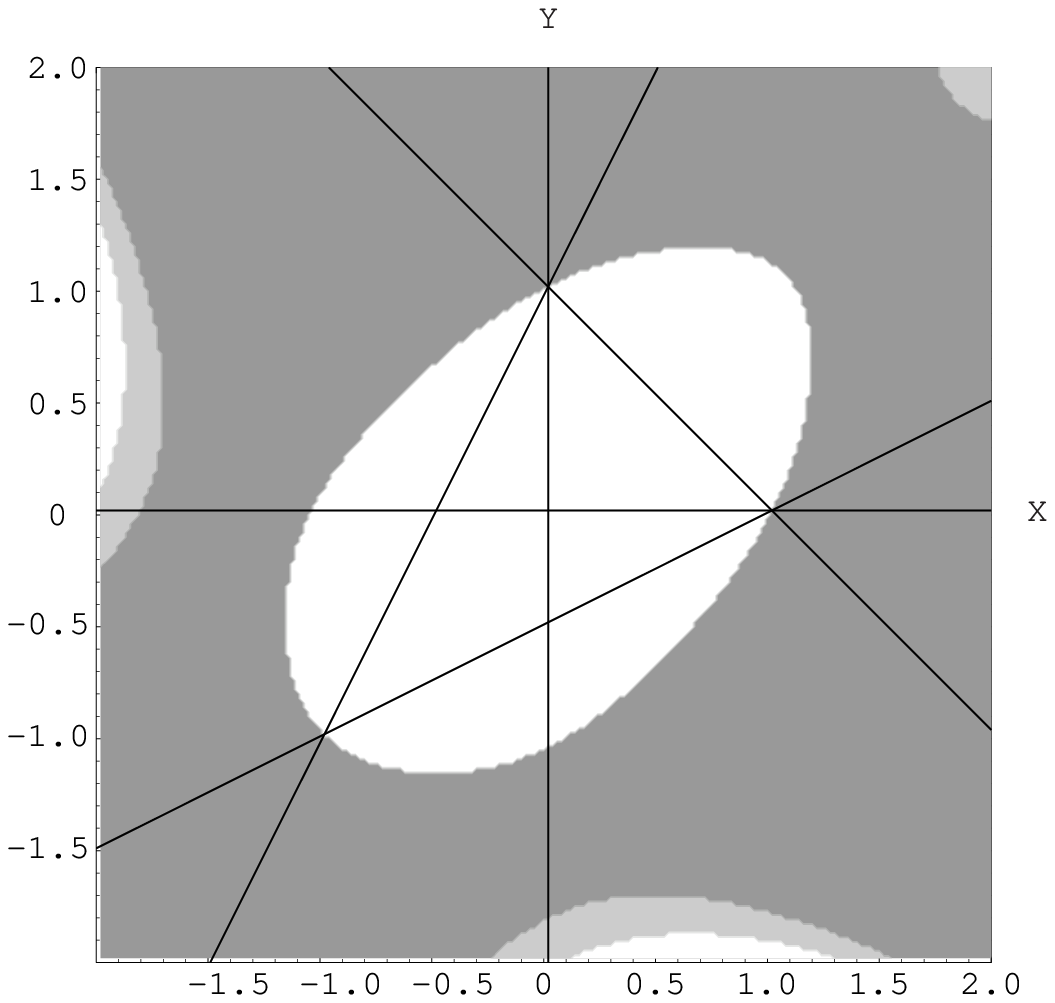}

$H_{\mbox{\scriptsize avr}}=0.2$, \hspace{2mm}$\sigma=2.5$\hspace{3cm}{}
\epsfxsize=6cm
\epsfysize=6cm
\epsffile{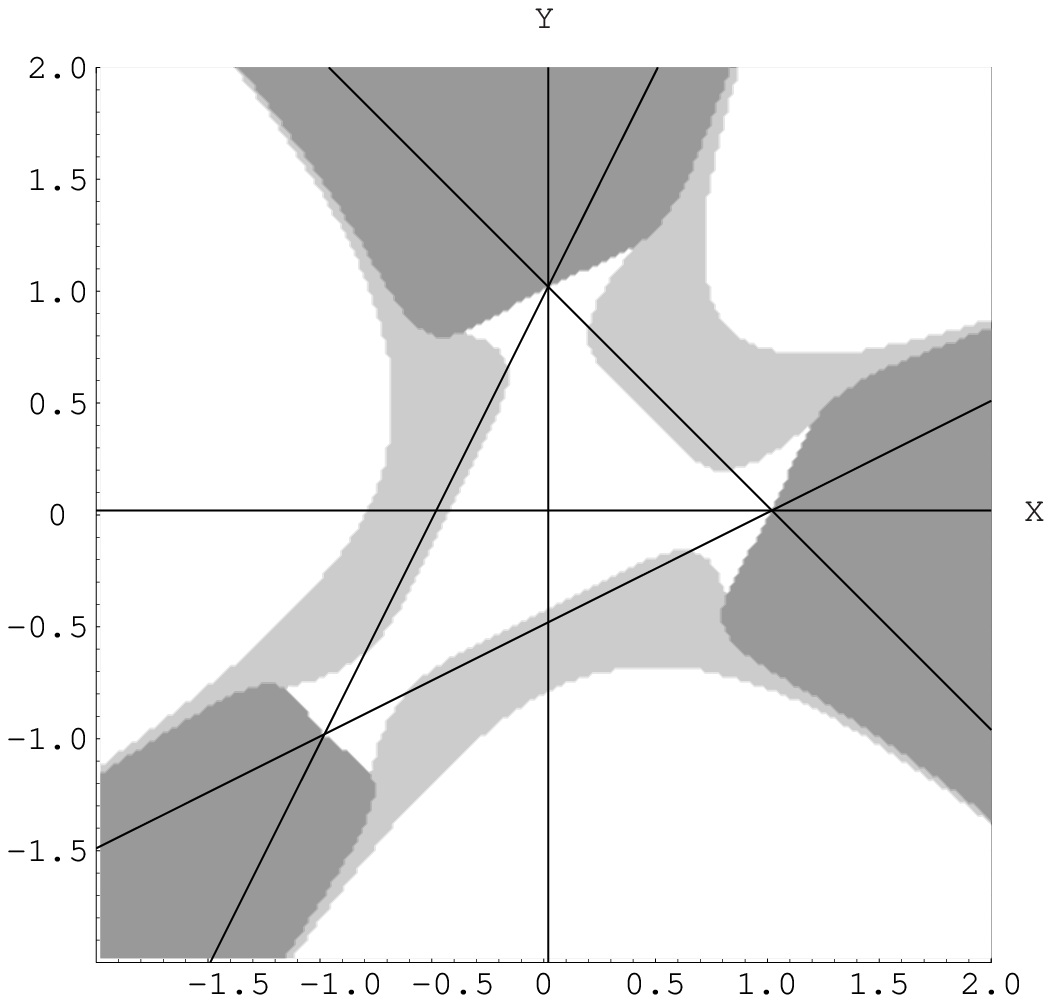}

$H_{\mbox{\scriptsize avr}}=0.3$, \hspace{2mm}$\sigma=2.5$\hspace{3cm}{}
\epsfxsize=6cm
\epsfysize=6cm
\epsffile{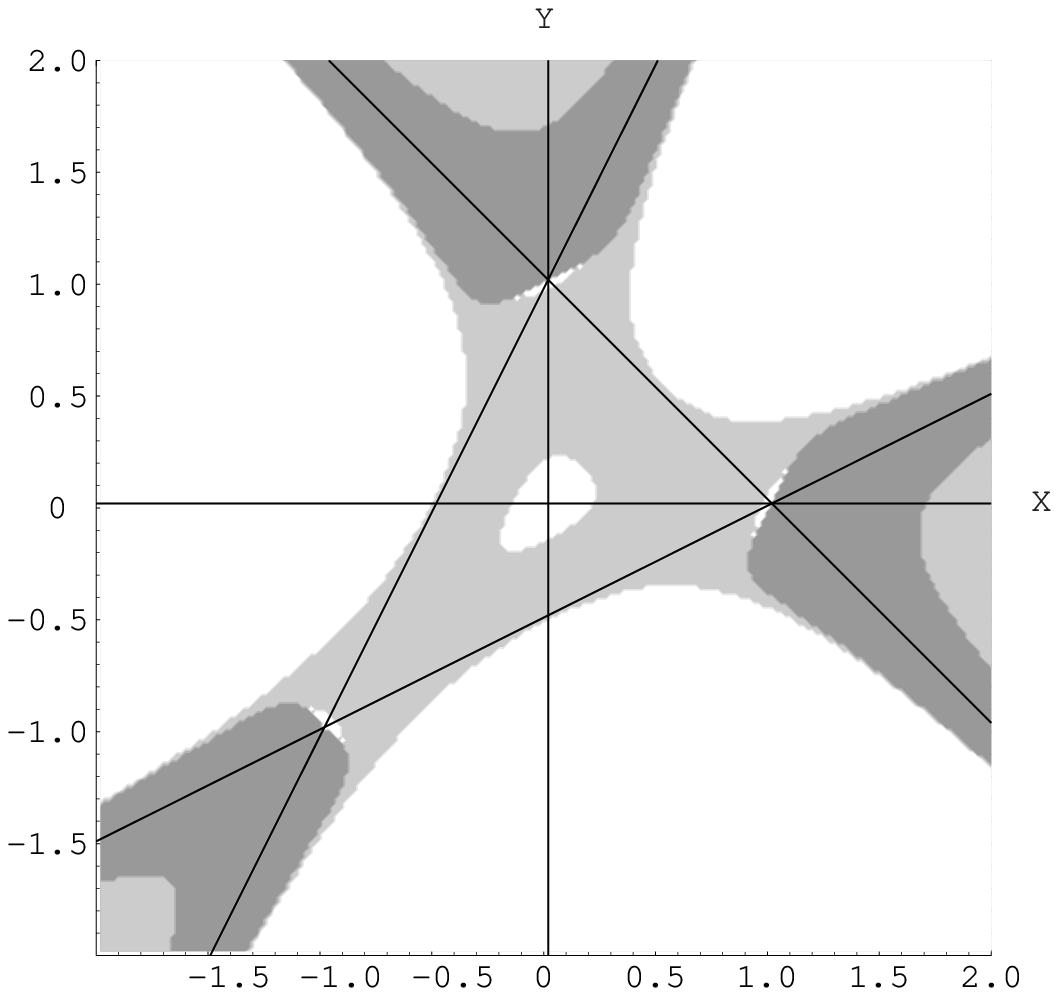}
\caption{Constraint and $\Delta$ on the $\sigma=2.5$ section of the $X$-$Y$ 
plane. $H_{\mbox{\scriptsize avr}}$ is 0.1, 0.2, 0.3 from above.
$\Delta>0$, $\Delta<0$, and excluded regions are indicated by white, 
light-shaded, and dark-shaded areas, respectively. }
\end{figure}
\end{center}
\onecolumn
\begin{figure}
$H_{\mbox{\scriptsize avr}}=0.01$, \hspace{2mm}$\sigma=2.5$\hspace{12mm}
$H_{\mbox{\scriptsize avr}}=0.05$, \hspace{2mm}$\sigma=2.5$\hspace{12mm}
$H_{\mbox{\scriptsize avr}}=0.1$, \hspace{2mm}$\sigma=2.5$\hspace{12mm}
\epsfxsize=6cm
\epsfysize=6cm
\epsffile{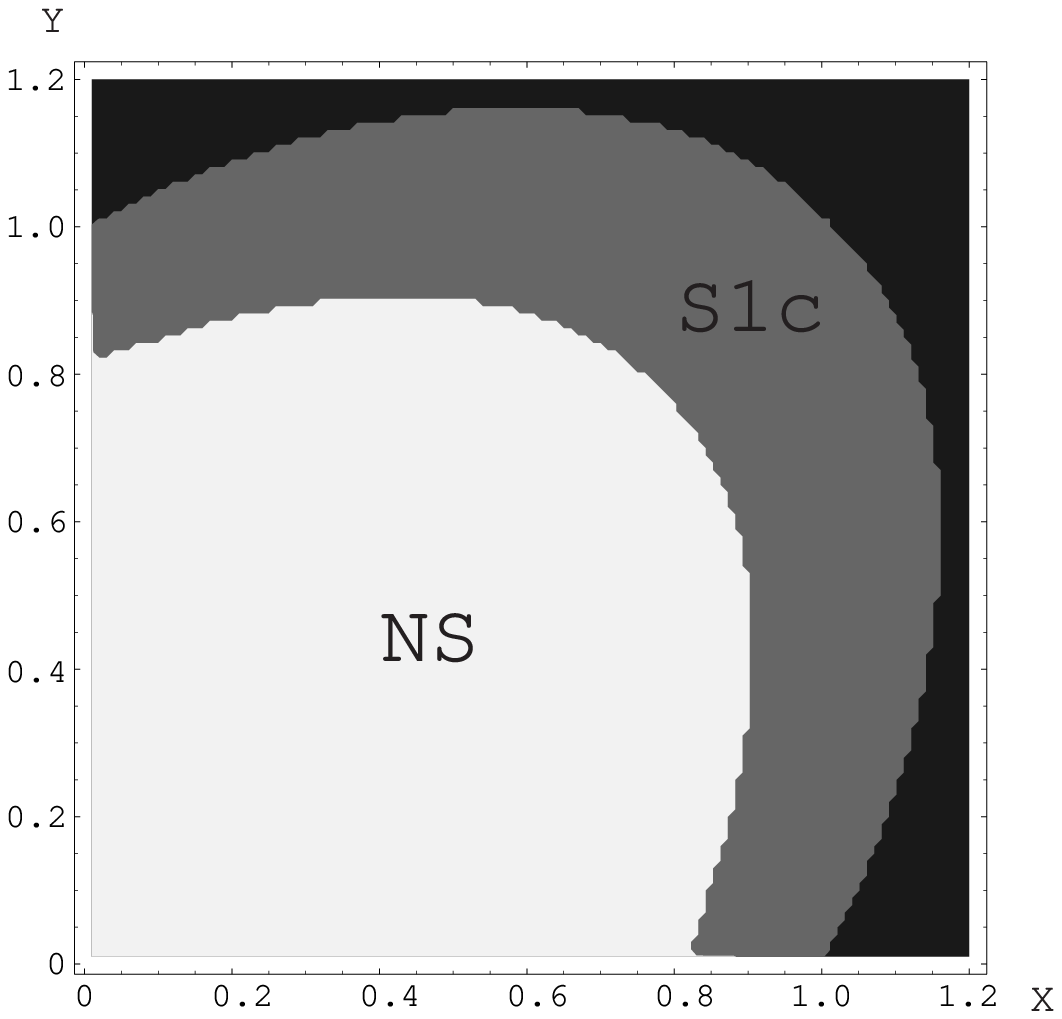}
\epsfxsize=6cm
\epsfysize=6cm
\epsffile{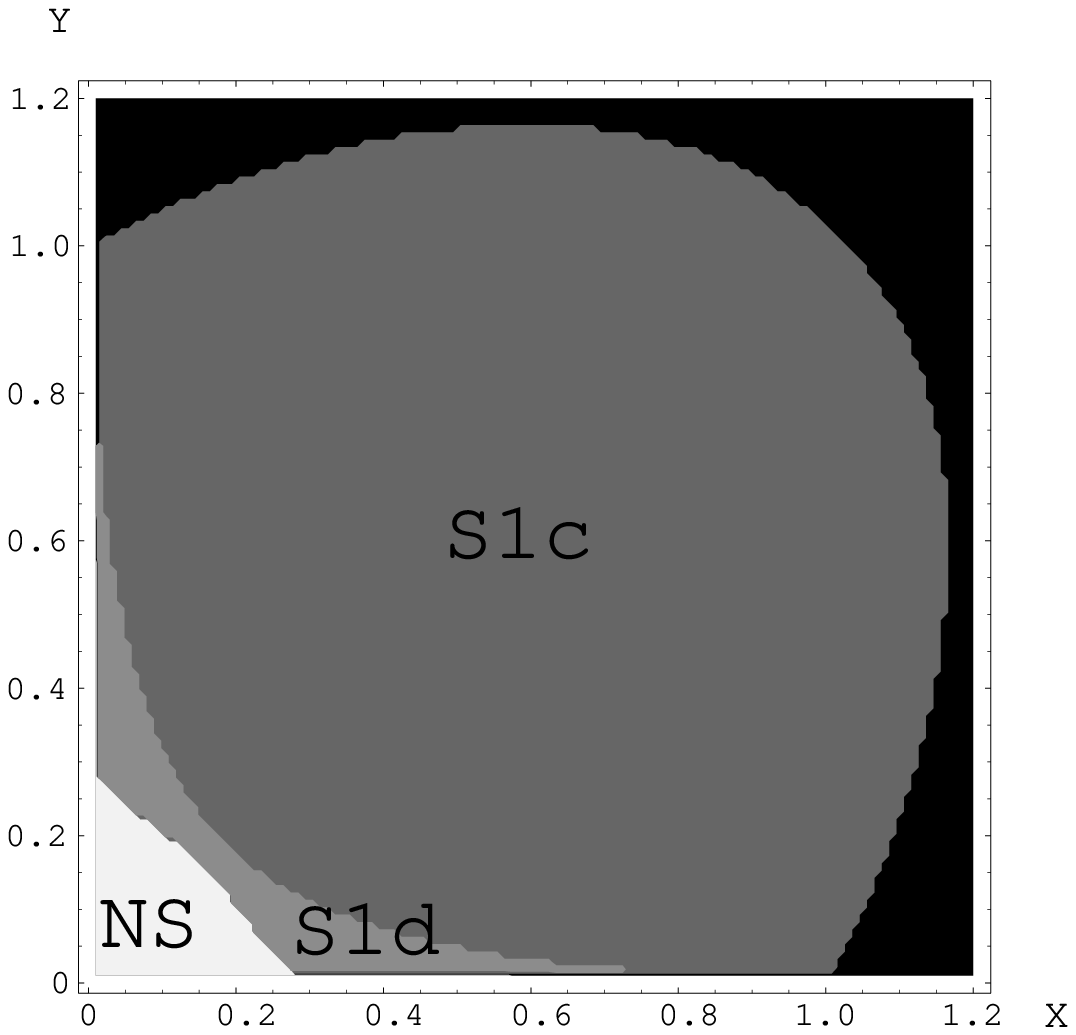}
\epsfxsize=6cm
\epsfysize=6cm
\epsffile{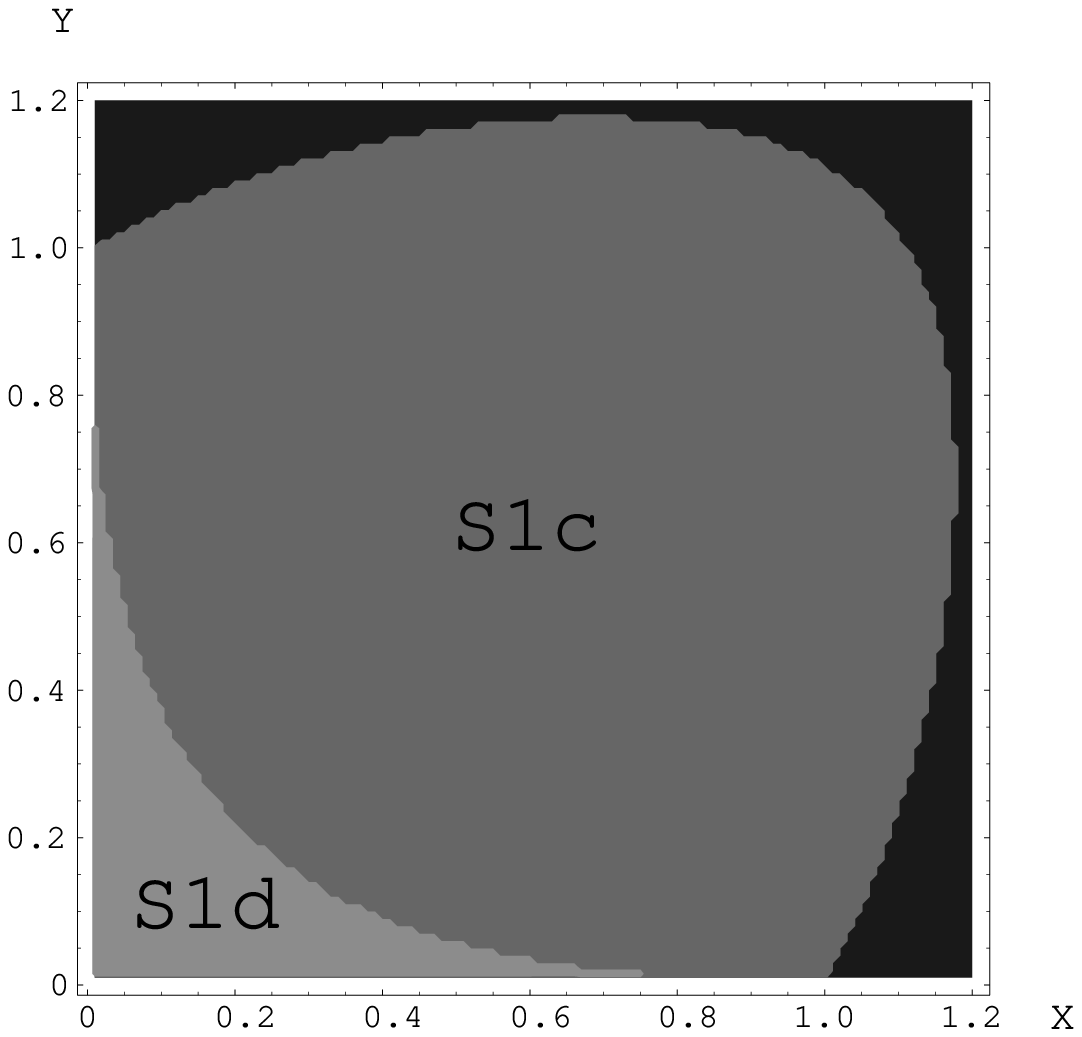}
\caption{Solutions through the $\sigma=2.5$ cross section. 
$H_{\mbox{\scriptsize avr}}$ is 
0.01, 0.05, 0.1 from the left. NS means nonsingular, which leads to an 
expanding universe in the past asymptotic region. 
S1 means it leads to a singularity where $\Delta\rightarrow 0$. 
S1c is the solution whose behavior near such a singularity is $p>0$, $q>0$, 
$r<0$, while S1d behaves as $p>0$, $q>0$, $r>0$, near the singularity.}
\end{figure}

\begin{figure}
\hspace{10mm}NS\hspace{75mm}S1c
\epsfxsize=8cm
\epsfysize=5cm
\epsffile{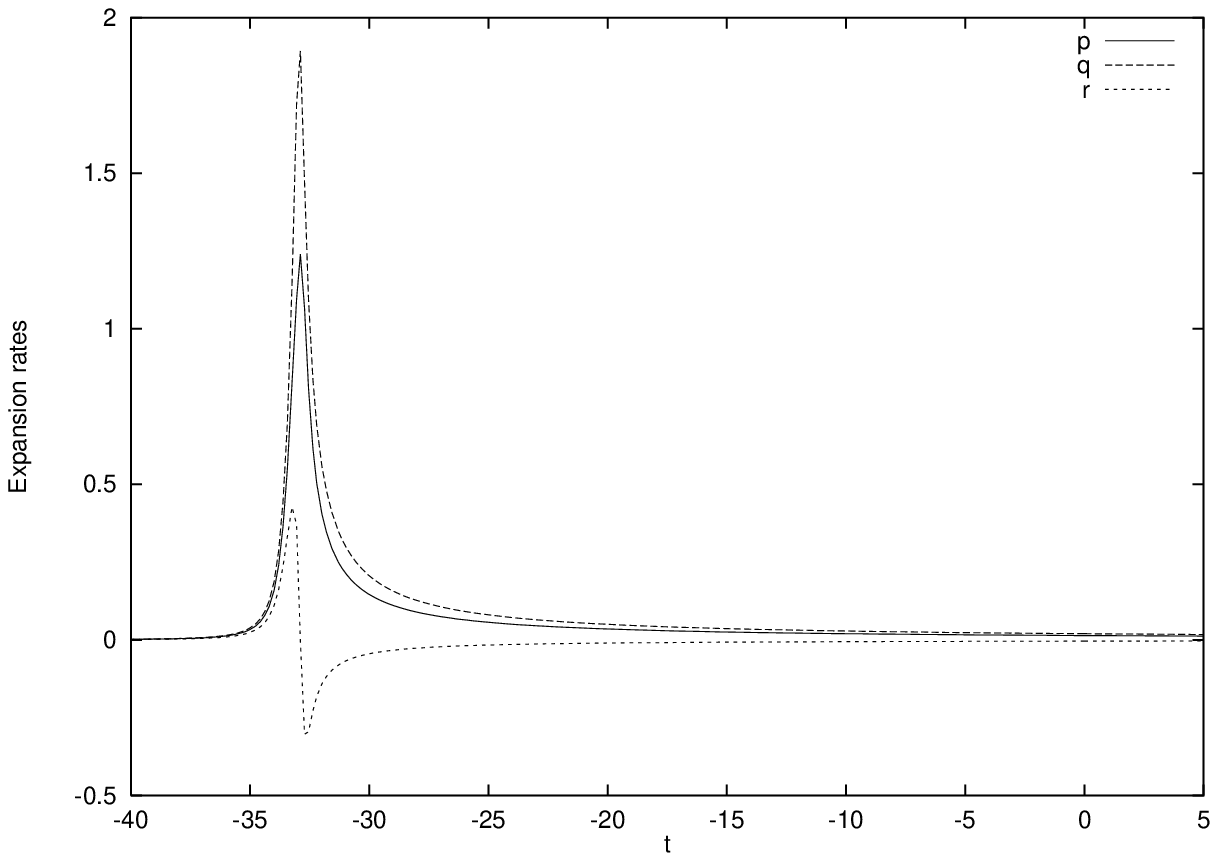}
\epsfxsize=8cm
\epsfysize=5cm
\epsffile{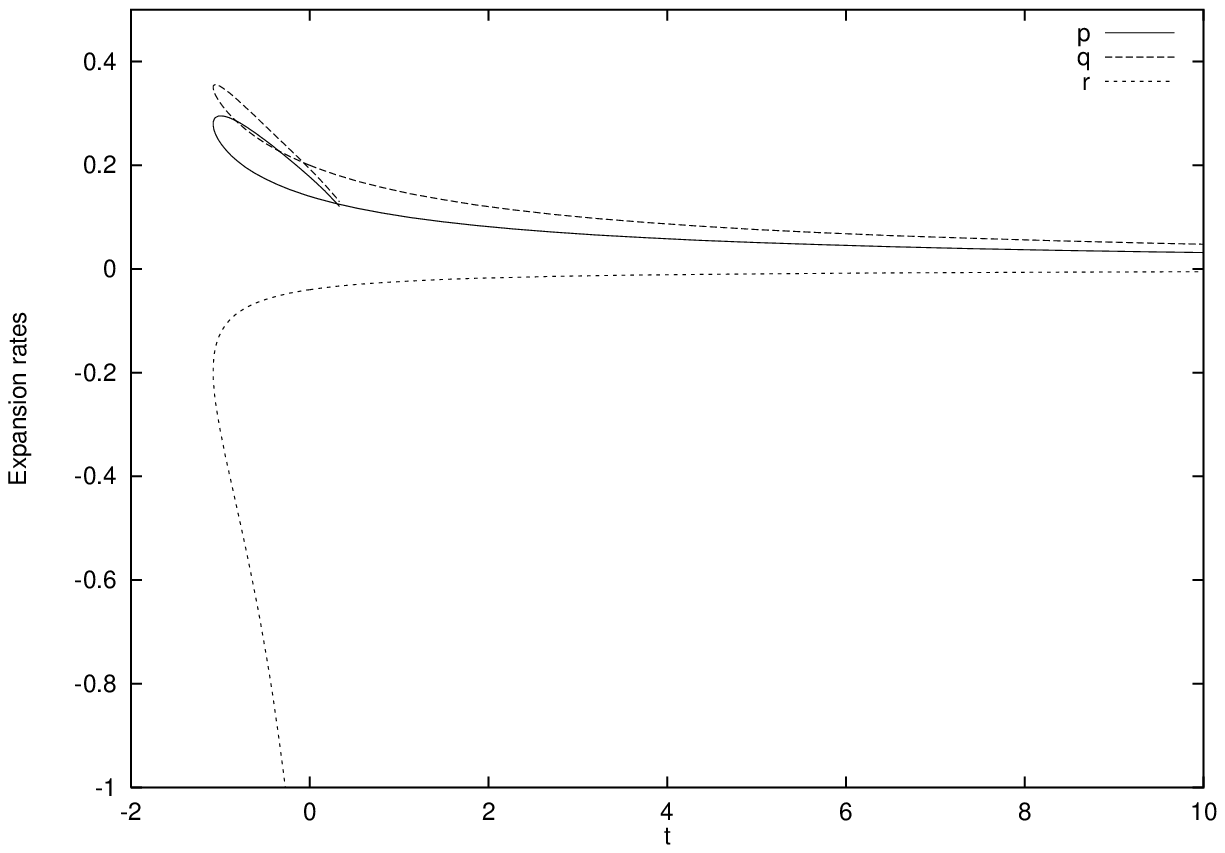}

\hspace{10mm}S1d\hspace{75mm}Isotropic
\epsfxsize=8cm
\epsfysize=5cm
\epsffile{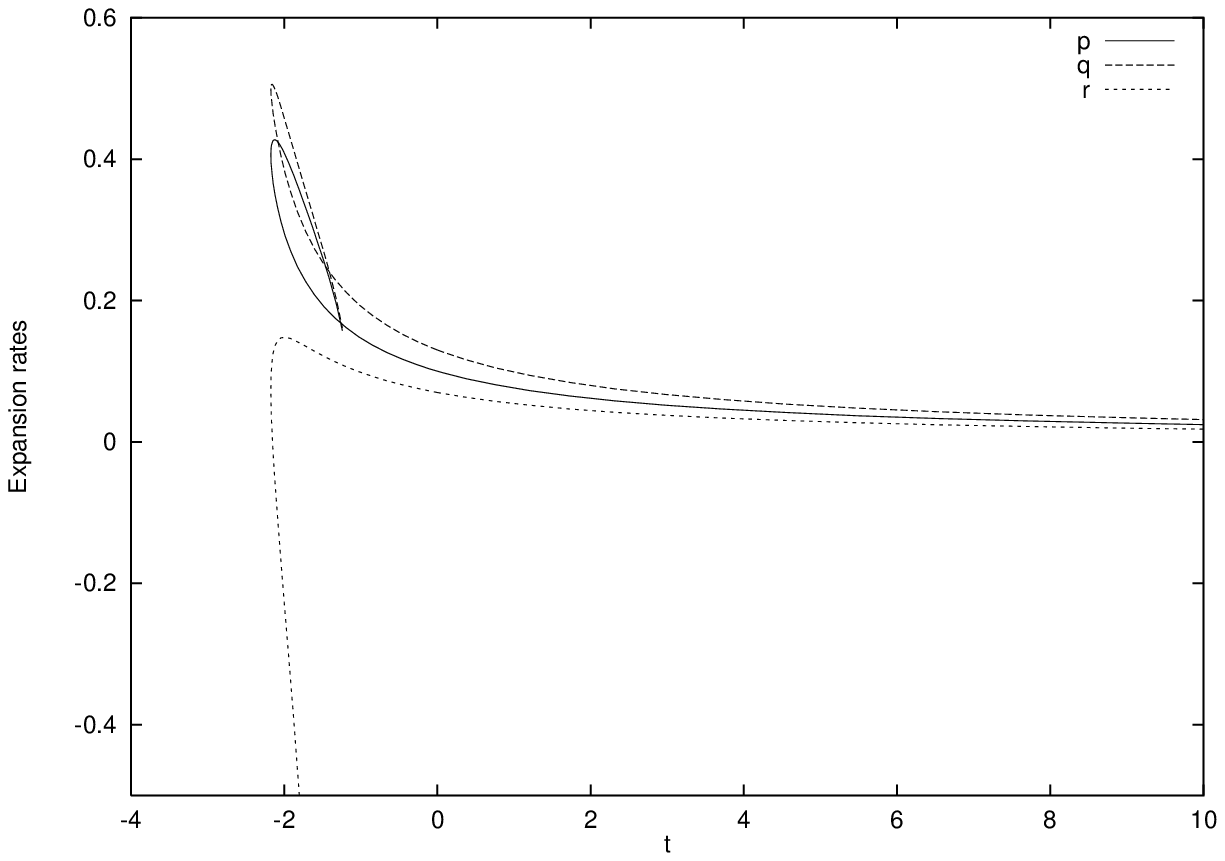}
\epsfxsize=8cm
\epsfysize=5cm
\epsffile{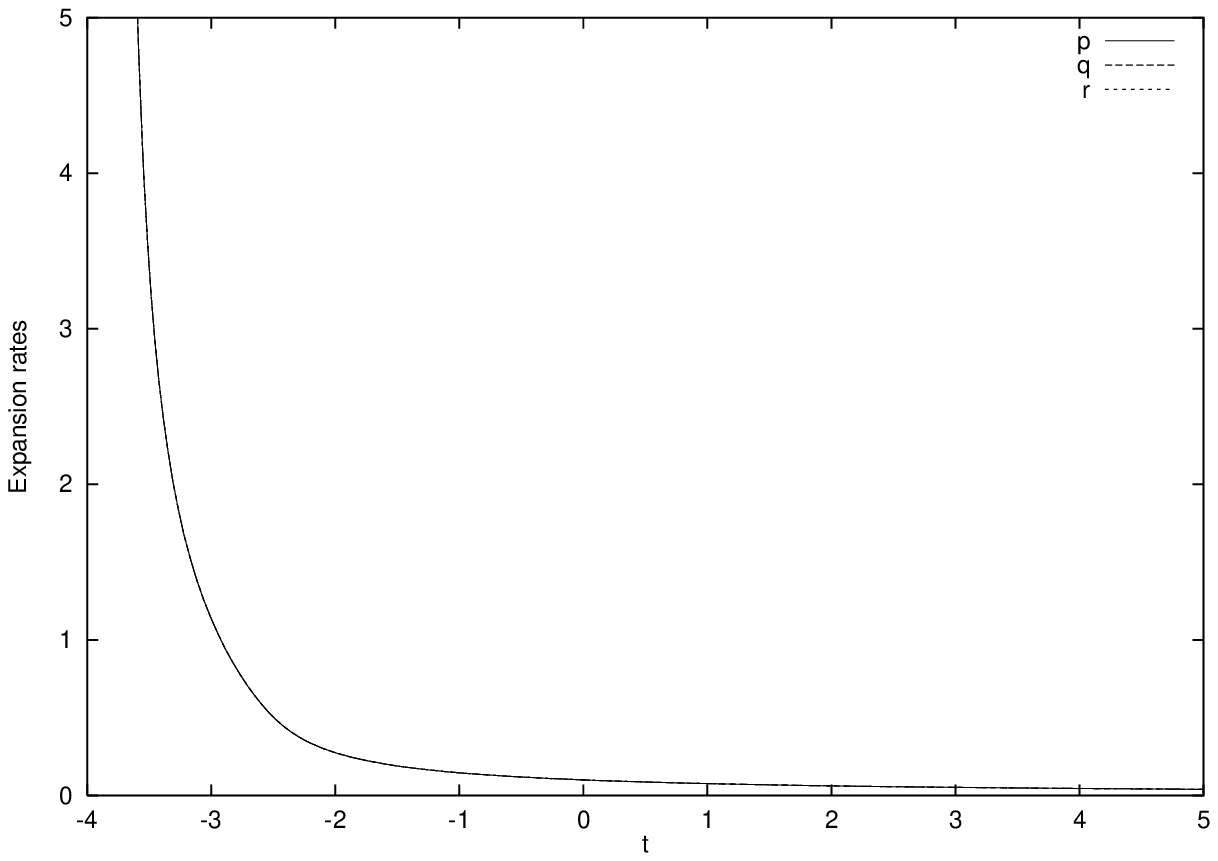}
\caption{Behavior of solutions appearing in Fig. 7. $t=0$ is the 
time when $\sigma=2.5$. The initial values are chosen on the $\sigma=2.5$
plane as
$(H_{\mbox{\scriptsize avr}},X,Y)=(0.01,0.6,0.8)$, $(0.1,0.6,0.8)$, 
$(0.1,0.1,0.2)$, and
$(0.1,0.0,0.0)$ for NS, S1c, S1d, and isotropic examples of the solution, 
respectively.}
\end{figure}

\end{document}